\documentclass[a4paper,fleqn,usenatbib]{mnras}

\usepackage[T1]{fontenc}
\usepackage{ae,aecompl}

\usepackage{graphicx}	
\usepackage{longtable}
\usepackage{lscape}
\usepackage{amsmath}	
\usepackage{amssymb}	

\title{An updated catalogue of giant radio sources}

\author[A. Ku\'zmicz et al.]{
A. Ku\'zmicz,$^{1,2,3}$\thanks{E-mail: cygnus@oa.uj.edu.pl}
M. Jamrozy,$^{2}$
K. Bronarska$^{}$
K. Janda-Boczar$^{2}$
D.~J. Saikia$^{4,5,6}$
\\
$^{1}$Center for Theoretical Physics, Polish Academy of Sciences, Al. Lotnik\'ow 32/46, 02-668 Warsaw, Poland\\
$^{2}$Astronomical Observatory, Jagiellonian University, ul. Orla 171, 30-244 Krakow, Poland\\
$^{3}$Queen Jadwiga Astronomical Observatory in Rzepiennik Biskupi, 33-163 Rzepiennik Strzy\.zewski, Poland\\
$^{4}$Inter-University Centre for Astronomy and Astrophysics (IUCAA), Ganeshkhind, Pune 411 007, India\\
$^{5}$NCRA, TIFR, Post Bag 3, Ganeshkhind, Pune 411 007, India\\
$^{6}$Cotton University, Panbazar, Guwahati 781 001, India}

\date{Accepted August 10, 2018}

\pubyear{2018}

\begin{document}
\label{firstpage}
\pagerange{\pageref{firstpage}--\pageref{lastpage}}
\maketitle

\begin{abstract}
We present a catalogue of 349 giant radio sources (GRSs including both galaxies and quasars). The database contains all giants known to date from the literature. These GRSs cover the redshift range of 0.016$<$z$<$3.22 and include radio sources of projected linear sizes larger than 0.7~Mpc which extend up to 4.7~Mpc. We provide the principal parameters (i.e. exact position of the host in the sky, redshift, angular and projected linear size, red optical magnitude, radio morphology type, total radio flux density and luminosity) for all the sources as well as characteristics of the sample. Based on the distribution of GRSs in the sky we identify regions where there is a paucity of giants, so that future surveys for this type of objects could concentrate primarily in these fields. From the analysis presented here, we estimate a lower limit for the expected number of GRSs as about 2000, for the resolution and sensitivity limits of FIRST, NRAO VLA Sky Survey and Sloan Digital Sky Survey surveys. Compared with earlier compilations, there is a significant increase in the number of large giants with sizes $>$ 2 Mpc as well as those at high redshifts with z$>$1. We discuss aspects of their evolution and suggest that these are consistent with evolutionary models.
\end{abstract}

\begin{keywords}
galaxies: active -- galaxies: nuclei -- galaxies: structure -- quasars: general -- radio contiuum: galaxies
\end{keywords}



\section{Introduction}

The giant radio sources (GRSs) are defined as extragalactic radio sources, hosted by galaxies (giant radio galaxies; GRGs) or quasars (giant radio quasars; GRQs), for which the projected linear size of radio structure is larger than 0.7 Mpc{\footnote{Many earlier authors, assuming $H_{0}$=50 km/s/Mpc, have used a lower limit of 1 Mpc as the defining size for GRSs. For the currently accepted cosmological parameters as given above, a limiting size of $\sim$0.7 Mpc is appropriate.}} (assuming $H_{0}$=71 km/s/Mpc, $\rm \Omega_{M}$ = 0.27, $\rm \Omega_{vac}$ = 0.73). The first objects of this type to be found were 3C236 and DA240, identified by \citet{willis1974},  while the first sample of 53 GRSs compiled from the literature was published by \citet{ishwara1999}. Since that time, the number of these relatively rare sources have been growing in number due to the all-sky surveys and the increasing amount of sensitive and good-quality radio and optical data. Many research groups searched for giants over the entire sky using modern radio surveys like the Westerbork Northern Sky Survey (\citealt{rengelink1997}) at 0.3 GHz, the Sydney University Molonglo Sky Survey (SUMSS, \citealt{bock1999}) at 0.8 GHz, the NRAO VLA Sky Survey (NVSS, \citealt{condon1998}) and the Faint Images of the Radio Sky at Twenty-Centimeters (FIRST, \citealt{becker1995}) at 1.4 GHz. The more recent samples by \cite{schoenmakers2000}, \cite{saripalli2005}, \cite{lara2001}, \cite{machalski2001,machalski2006,machalski2007}, \cite{kuzmicz2012} as well as \cite{dabhade2017} and others have increased the number of known GRSs more than 6 times as compared to the sample of \cite{ishwara1999}.

However, GRSs may not be easily recognizable if the surface brightness of the lobes is low. This could be the case for the less luminous GRSs with large angular sizes. Also, at high redshifts, where the inverse-Compton losses against the cosmic microwave background radiation are large, extended radio sources could have weak radio lobes (e.g. \citealt{konar2004}). In these cases the hot-spots or lobes may not appear to be connected with the radio core, thereby making the detection of GRSs quite a challenging task. Due to all these reasons as well as the steep spectra of many GRSs, the new-generation radio interferometers such as the LOw-Frequency ARray (\citealt{haarlem2013}), the Murchison Widefield Array (MWA; \citealt{lonsdale2009}), the Square Kilometre Array (SKA; \citealt{carilli2004}) or upgrades of existing low-frequency telescopes such as the Giant Metrewave Radio Telescope (GMRT; \citealt{gupta2017}) should be of particular help in finding more of these objects.  

From previous studies, it is still unclear why just a small fraction of radio sources reach such large sizes. Nevertheless, owing to the observations conducted in the last decade, our knowledge of the nature of GRSs has progressed significantly. These investigations focused on the role of the properties of the intergalactic medium (IGM; \citealt{machalski2006}; \citealt{subrahmanyan2008}; \citealt{kuligowska2009}), the advanced age of the radio structures (e.g. \citealt{mack1998}; \citealt{machalski2009}), recurrent radio activity (e.g. \citealt{subrahmanyan1996}; \citealt{schoenmakers2000}; \citealt{saikia2009}; \citealt{machalski2011}), and specific properties of the central active galactic nuclei (AGNs; e.g. black hole mass and accretion rate, \citealt{kuzmicz2012}), as underlying reasons for their gigantic sizes. They revealed that GRSs are quite similar to radio sources with smaller radio structures, but significantly older. There is a trend for the spectral ages of radio galaxies to increase with linear size (e.g. \citealt{murgia1999}; \citealt{parma1999}; \citealt{murgia2003}; \citealt{jamrozy2008}). However, there also exist small-sized radio galaxies with large ages ($\sim$10$^8$yr). \citet{murgia2011} presented a study of five very aged radio galaxies with linear sizes of only about 100 kpc. Considering the IGM properties, it has been measured that in the vicinity of some GRSs the IGM density is lower (e.g. \citealt{schoenmakers2000b}; \citealt{lara2000}), while we need to understand why we do observe giants at higher redshifts (z$>$1) where the IGM density is actually higher than at present. Moreover, \cite{komberg2009} have shown that there is no correlation between the radio source size and the density of galaxies in the neighbourhood. However, we still lack an unequivocal explanation for the giant sizes of some of these radio sources. It is obvious that GRSs are very interesting objects worth intensive research. Investigations of their properties are necessary to fully understand the processes responsible for the formation and evolution of radio sources in general. 

Moreover, GRSs can  be useful for cosmological studies. They can help in determining the IGM properties at different redshifts. A number of studies carried out during the last several years show that GRSs are a very valuable tool for investigating the large-scale structure of the Universe. Their large sizes provide an opportunity to probe the distribution of the Warm-Hot Intergalactic Medium (WHIM) in filaments of the large-scale structure of the Universe (e.g. \citealt{malarecki2015, malarecki2013}; \citealt{pirya2012}; \citealt{peng2015}). These studies focus on searching interactions of radio lobes with the ambient medium revealed through asymmetries of radio structures and the distribution of neighbouring galaxies. It is believed that through these kinds of investigations it will be possible to find the `missing' baryons predicted by the Big Bang theory (\citealt{peng2015}).

Furthermore, the recent investigations of \cite{bassani2016} show that a large fraction of soft gamma-ray selected radio sources become GRSs. The all-sky observations of the INTErnational Gamma-Ray Astrophysics Laboratory (INTEGRAL; \citealt{winkler1994}) and Swift (\citealt{gehrels2004}) satellites reveal a large population of AGNs. The double-lobed radio sources are not too common in this group (about 7\%), yet it is very intriguing that a large fraction of them are giants (23\%). It may be the case that high-energy surveys could be more efficient in searching for new GRSs as compared to the radio surveys, where for example, in the well-studied 3CRR sample about 8\% of the sources are identified as GRSs.

Given the above background, it is clear that it is highly advisable to look for new giants, and compiling a large sample of GRSs will facilitate future research in this field. 

In this paper, we present a catalogue of GRSs known to date. We provide their principal parameters as well as characteristics of the sample. In this compilation of GRSs we do not consider objects such as cluster radio relics and/or radio halos that can also exceed our defining size of 700 kpc. The content of the paper is as follows. Section 2 describes details of the GRSs catalogue. In Sections 3 and 4 we analyse and discuss their distribution, physical parameters, and aspects of their evolution, while in Section 5 we present our concluding remarks.

\section{Catalogue}

For several years we have been browsing the literature, including survey results, and analysing data to search for radio galaxies of large linear sizes. We focused mostly on the existing compilations of giants, as well as published studies of individual sources of such types, but also examined well-studied samples such as 3CRR and compilations of structures of radio sources such as, for example, that by \cite{nilsson1998}. In effect we have compiled a list of all GRSs known to the end of 2017. The list and principal parameters of GRSs are presented in Table 1 which is arranged as follows: Col. 1 -- source name; Cols. 2 and 3 -- J2000.0 coordinates of the GRS host galaxy or quasar; Col. 4 -- optical identification (G -- galaxy or Q -- quasar); Col. 5 -- redshift; Col. 6 -- radio morphological type based on the Fanaroff-Riley classification scheme (see below); Col. 7 -- angular size in arcmin; Col. 8 -- projected linear size, D, in Mpc; Col. 9 -- {\it r} band optical aperture magnitude; Col. 10  -- total flux-density at 1.4 or 1.388 or 0.843 or 0.325 GHz in units of mJy; Col. 11 -- error in total flux-density; Col. 12 -- 1.4 GHz total radio luminosity in W/Hz; Col. 13 -- References. 

The redshifts enclosed within parentheses correspond to the photometric redshifts taken either from the literature or from the Sloan Digital Sky Survey (SDSS; \citealt{Albareti2017}). The 1.4 GHz flux-densities were measured mostly on the maps of the NVSS using the AIPS{\footnote{www.aips.nrao.edu/index.shtml}} software and particularly the task `tvstat'. During the measurements we have made efforts to exclude unrelated (foreground/background) sources superposed on the extent of some GRSs. For some sources with declination below $-$40$\rm^{o}$, we used the flux-density data at 1.388 GHz taken from \citet{saripalli2012}. They are marked as {\it `a'} in the Table 1. The SUMSS flux-density measurements are marked as {\it `b'}, the measurements at 0.843 GHz given by \cite{saripalli2005} are marked as letter {\it `c'} and the measurements at 0.325 GHz given by \cite{sebastian2018} are marked as letter {\it `d'}. In order to determine the flux-density error, we used the formula given by \cite{klein2003}:
\begin{equation}
\delta \rm S=\sqrt{(S\cdot0.03)^2+(\sigma\cdot\sqrt{\frac{\Omega_{int}}{\Omega_{beam}}})^2},
\end{equation}
where `S' is the measured total flux-density in mJy, the 0.03 value corresponds to the assumed 3\% calibration error; `$\sigma$' is the r.m.s. noise measured around the source, `$\Omega_{int}$' is the integration area, and `$\Omega_{beam}$' is the beam's solid angle. The radio morphological type (FRI, FRII; \citealt{fanaroff1974}) was predominantly taken from the literature but for newly classified sources we judged it by ourselves using high-resolution radio maps. To determine the total radio luminosity, we used the formula given by \citet{brown2001} and assumed for all sources a mean spectral index value as $\alpha =-0.8$ (following e.g. \citealt{nilsson1998}). For sources with only 0.843 or 0.325 GHz flux density measurements available we determined the 1.4 GHz flux density using the above given spectral index value.

For classical FRII sources, which constitute about 90 per cent of the sources in our list, we measure the angular extent as the size between the hotspots. Our measurements and those given by other authors are similar for these types of sources. In cases when we do not have an image of good enough angular resolution for a source, we adopt the angular size as was given in the literature. Controversies regarding angular size measurement appear mostly in the case of FRI (and hybrid FRI/FRII) sources. For such objects we try to estimate the angular size from a map/publication at which the source shows a maximal angular radio extent and its size was taken as the distance between the opposite edges. Something similar occures for FRII sources (e.g. J0116$-$4722, J1548$-$3216) which show a double-double radio structure (without clear hotspots in the outer lobes) or those that have at least a clear extension beyond the hotspot(s)  (e.g. DA240). In the case of sources which have a very curved morphology (e.g. wide-angle-tail, head tail radio galaxies) their angular extent was measured along the source ridge, and were not, as in the case of most giants, taken as the shortest distant between opposite edges.

The optical {\it r}-band aperture magnitudes are taken from the PanSTARRS data archive (\citealt{flewelling2016}) which has surveyed the sky north of $\delta$=$-$30$\rm^{o}$ in five photometric bands ({\it grizy}). For some objects south of $\delta$=$-$30$\rm^{o}$ we have taken {\it r}-band magnitudes from \cite{saripalli2012} and marked them as `r' in Table 1.

The references given in Table 1 are mostly related to papers where a radio source has been recognised as a giant for the first time, or to papers where some important parameters characterising a particular object (e.g. redshift, radio map) were published. In Table 1 we include only the confirmed GRSs, though there are many more GRS candidates with redshifts still to be determined.

The final catalogue includes 349 sources, of which 280 are galaxies, 68 are quasars, and 1 is of uncertain optical identification, covering the redshift range of 0.016$<$z$<$3.22. In the sample, there are 46 objects with photometric estimation of redshifts. Redshifts larger than 1 were found for 22 GRSs based on spectroscopic measurements and for 6 GRSs based on photometric estimations. The most distant GRSs are J1145$-$0033 with redshift z=2.055 (\citealt{kuzmicz2011}) and J1235$+$3925 (i.e. 4C39.37) with redshift z=3.22 (\citealt{mack2005}). However, the extent of the radio protrusions of the latter object is not certain, since \cite{mack2005} mentioned that ``it cannot be excluded that these are artefacts caused by an imperfect amplitude calibration''. Furthermore, 4C39.37 is the highest luminosity GRS in our sample. Although its structure needs to be confirmed, it is consistent with the extrapolation of the log P -- z relation (Figure \ref{zp}) appearing as the most distant and luminous object.

More than half of the GRSs have projected linear sizes larger than 1 Mpc. Four of them have extremely large sizes: J1006$+$3454 (4.23 Mpc), J0931$+$3204 (4.29 Mpc), J1234$+$5318 (4.44 Mpc), J1420$-$0545 (4.69 Mpc). The last one is still the largest GRS known to date (\citealt{machalski2008}). Most of the catalogued sources have an FRII radio morphology. Only 20 sources reveal an FRI structure and 16 are classified as hybrid FRI/FRII sources.

This sample of GRSs as a whole is quite heterogeneous, including objects from studies in which sources were not selected in any systematic manner. Moreover, it is restricted by the selection effects related to sensitivity of radio and optical surveys. However, it may be possible to construct statistically complete subsamples in restricted areas of the sky that have been imaged uniformly by surveys such as NVSS/FIRST and optical surveys such as SDSS. 

\section{Sky coverage}

\begin{figure*}
\centering
    \includegraphics[width=1.99\columnwidth]{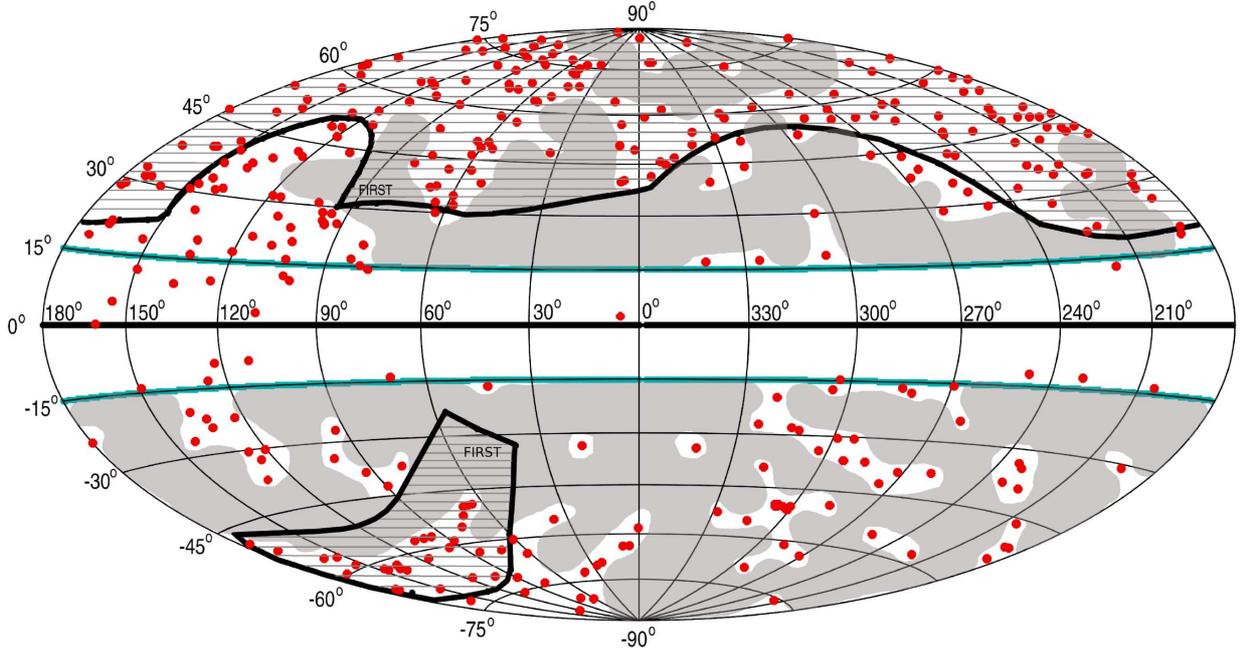}
\caption{Distribution of GRSs in the plane of the sky in Galactic coordinates. The blue lines are plotted at Galactic latitudes b=$\pm$15$\degr$ denoting the larger Galactic extinction regions. In gray we coloured the regions outside the Galactic plane where no GRSs have been found. The sky area that is covered by the FIRST survey is framed by a thick black curve and in addition the area is hatched.}
\label{gal}
\end{figure*}

\begin{figure*}
\centering
    \includegraphics[width=1.99\columnwidth]{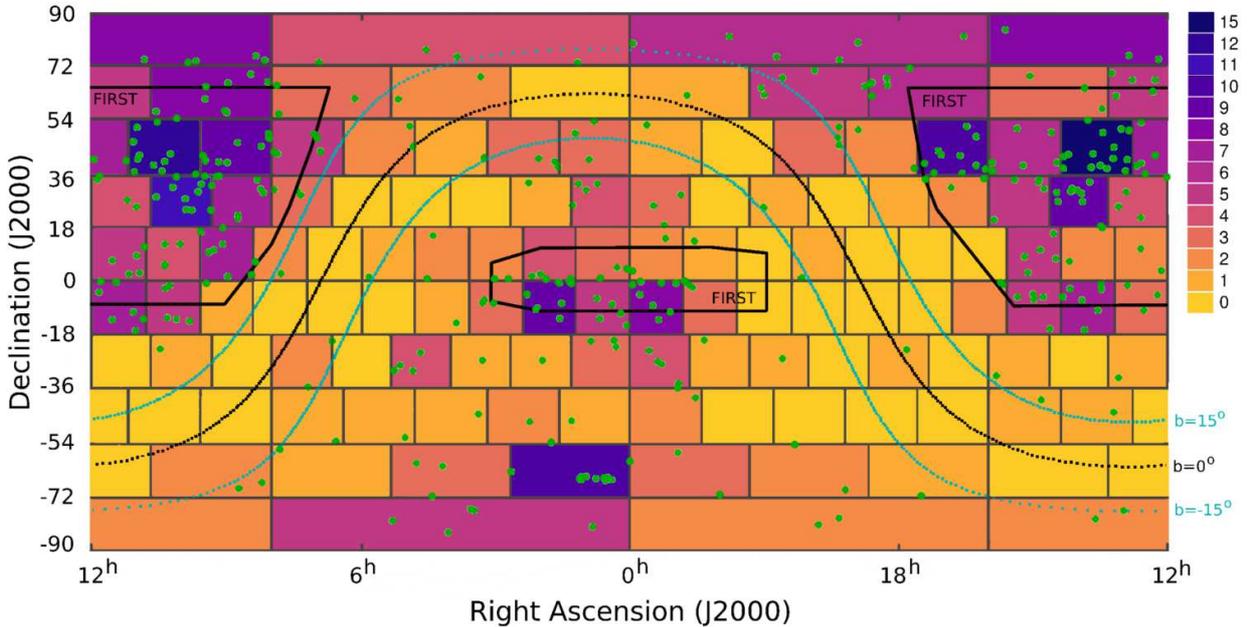}
\caption{The distribution of GRSs in the plane of the sky in equatorial coordinates. The blue dotted lines correspond to the Galactic latitudes b=$\pm$15$\degr$ and the black dotted line corresponds to the Galactic equator. The different colours denote the number of giants counted inside a particular rectangle and every rectangle covers a the same area of 317.33 deg$^2$ (\protect\citealt{malkin2016}).}
\label{eq}
\end{figure*}

In Figures \ref{gal} and \ref{eq} we plot the distributions of GRSs from our sample both in the Galactic and equatorial coordinates. It can be seen that it is not homogeneous and that a large fraction of known giants are seen in the northern hemisphere. On the sky maps there are regions where a lot of giants can be observed and regions without any recognized source (areas marked in gray in Figure \ref{gal}). In the strip along the Galactic equator ($|b|$$\lesssim$15$\degr$), the optical identification of GRSs can be difficult due to higher Galactic extinction. Therefore, infrared surveys (e.g. the Two Micron All-Sky Survey; 2MASS; \cite{skrutskie2006} or the Wide-field Infrared Survey Explorer (WISE; \cite{wright2010}) can be helpful for identifying host galaxies in this region. Moreover, there are a number of extended individual radio structures inside the Milky Way Galaxy that may confuse some of the GRSs' radio structure. The regions devoid of GRSs away from the Galactic plane are mostly due to incomplete coverage of both radio and optical surveys. Therefore, it would be good if future optical and radio survey efforts focus primarily on these `empty' fields for identifications of new GRSs. 

The largest number of giants is recognised in the regions that are covered by the FIRST radio survey, along with the availability of optical data (e.g. SDSS). As can be seen in Figure \ref{eq}, where the surface density of giants in the sky plane is depicted by different colours, the densest region is located in the area defined by 12.8$\rm^h$$<\alpha<$14.4$\rm^h$ and 36$\rm^{o}$$<\delta<$54$\rm^{o}$. There are 15 giants in this area, of which 12 are identified with galaxies and the remaining are identified with quasars. Similar numbers of GRSs are expected in other regions observed to similar radio and optical sensitivity limits. Extrapolating this density over the whole sky (4$\pi$ sr) and assuming a homogeneous distribution of GRSs, the total number of giants should be about 2000. This estimate does not take into account objects with low surface brightness that remain undetectable in the NVSS survey. Also, since most of our sources have been selected from low-frequency surveys (less than about 1.4 GHz), core-dominated GRSs are also likely to be under-represented in the sample.
Therefore, this is just a lower limit to the expected number of GRSs. Furthermore, the regions with the largest numbers of giants coincide with coverage of the FIRST survey, where better radio map resolution allowed for the identification of radio cores associated with host galaxies. However, it should be stressed that apart from regions that are not covered by both radio and optical surveys, there are also relatively large regions where FIRST and SDSS data are available but they still show a low  number of GRGs. For example, such regions are placed near the North Galactic pole (12$\rm^h$$<\alpha<$14.4$\rm^h$, 0$\rm^{o}$$<\delta<$18$\rm^{o}$ with density of only 2 GRSs over 317.33 deg$^2$) and in an area bounded by 21.8$\rm^h$$<\alpha<$3.6$\rm^h$, 0$\rm^{o}$$<\delta<$18$\rm^{o}$ (with an average density of 2.6 GRSs over 317.33 deg$^2$ ). The deficiency of known GRSs in those particular regions is not because of a lack of radio and optical data but due to a lack of conducting any systematic surveys for giants.


\section{Discussion}
\subsection{Physical parameters of the GRSs}
\begin{figure}
\centering
\includegraphics[width=0.99\columnwidth]{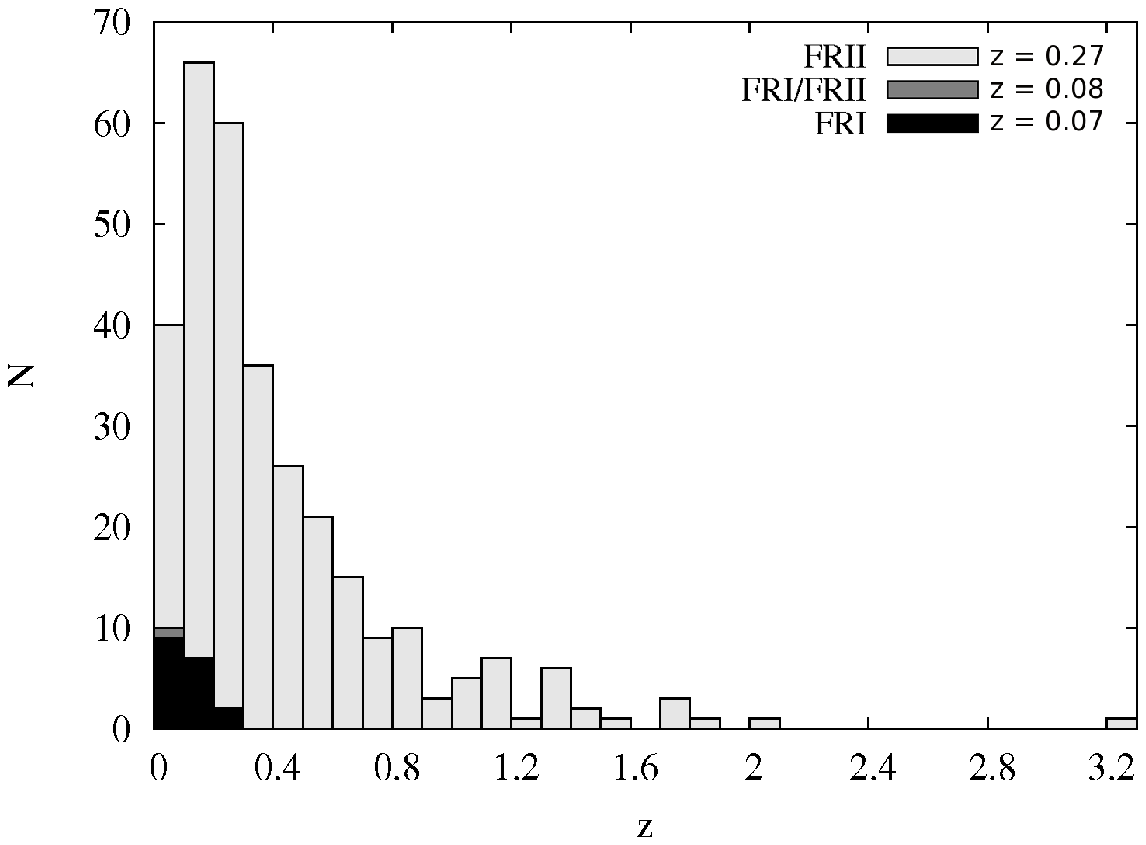}\\
\includegraphics[width=0.99\columnwidth]{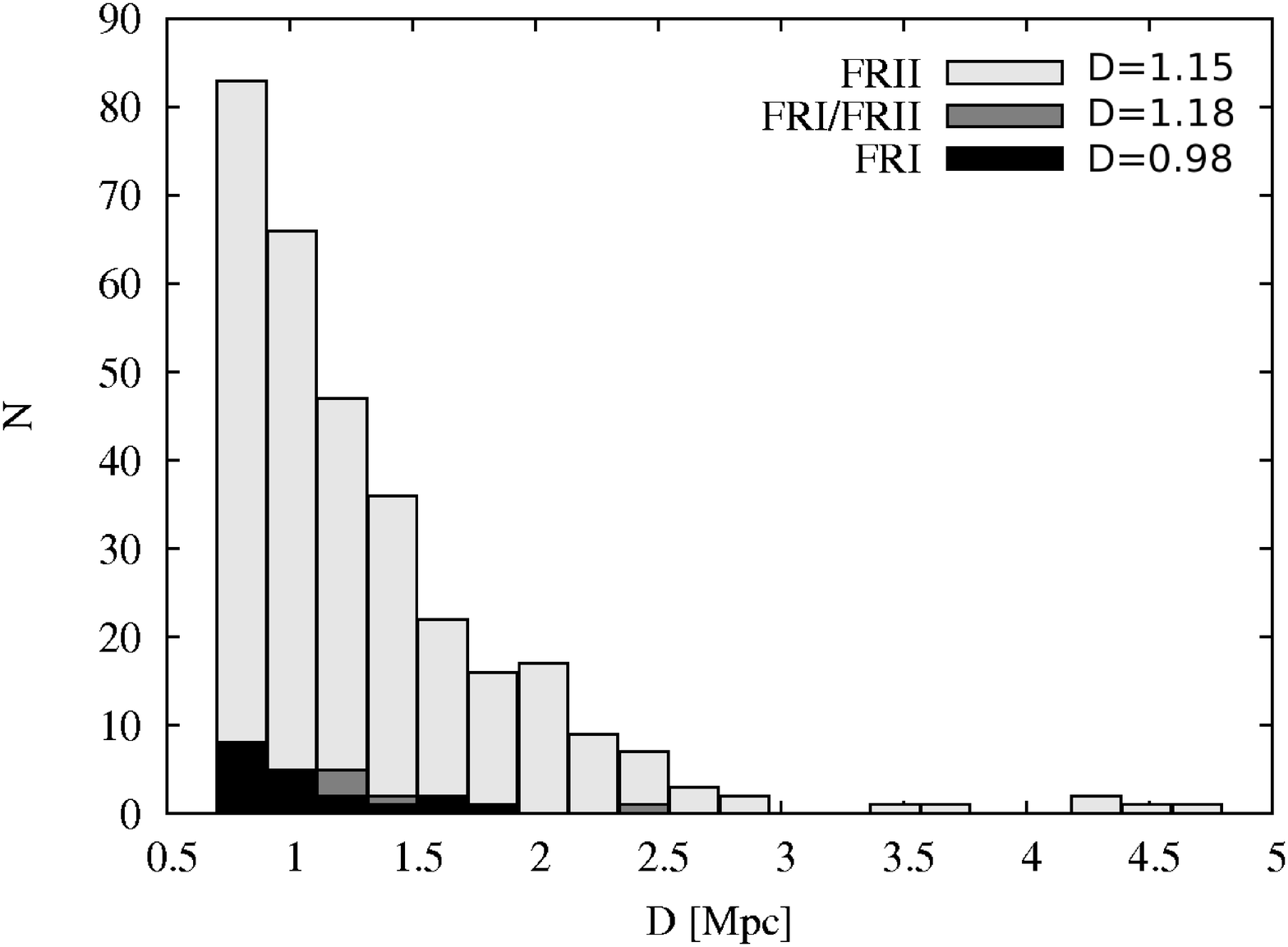}\\
\includegraphics[width=0.99\columnwidth]{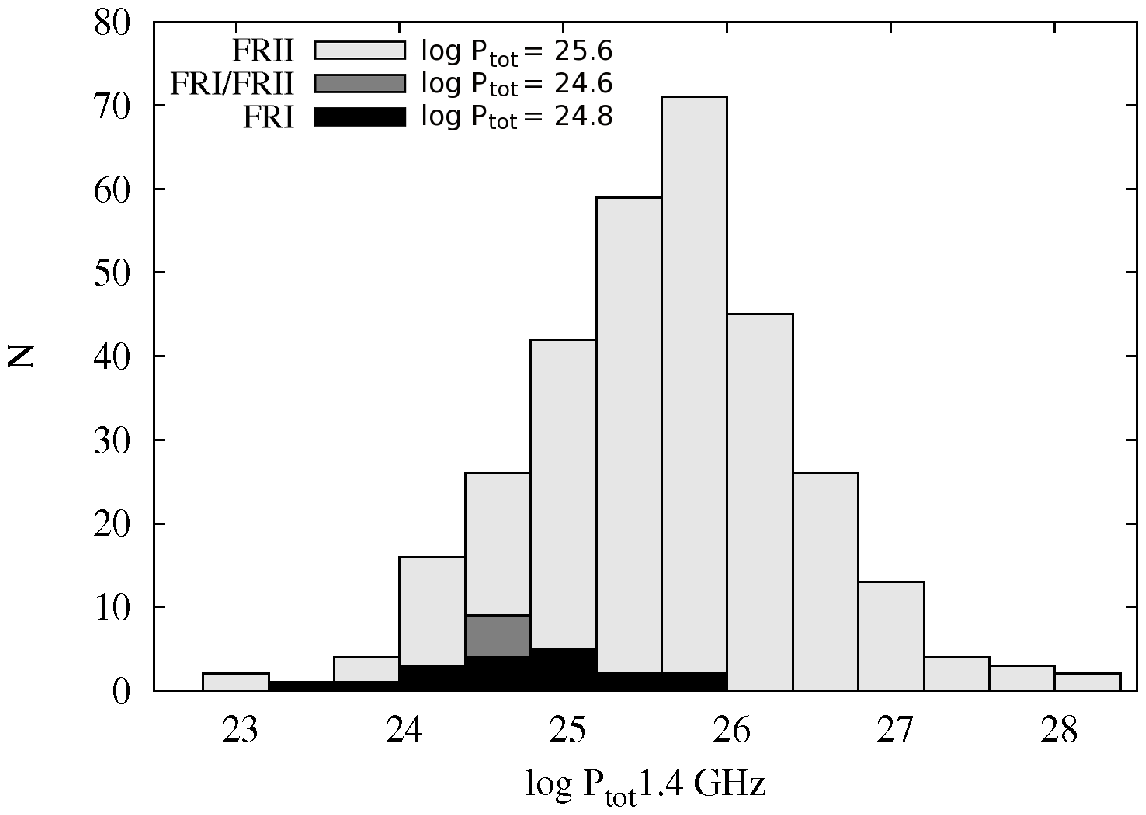}\\
\caption{Distributions of redshift (top panel), projected linear size (middle panel), and 1.4 GHz total radio luminosity (bottom panel) for GRSs. The number of sources N is put on the vertical axis. The values of redshift, projected linear size, and total radio luminosity given on the panels denote the median values for each morphological class of radio sources. }
\label{rozklad}
\end{figure}

In Figure \ref{rozklad}, we present distributions of redshifts, projected linear sizes and 1.4 GHz total radio luminosities for the GRSs catalogued in this paper. The linear sizes range from the cut-off value of 700 kpc to 4.69 Mpc, with a median value of 1.14 Mpc. In the earlier compilation by \cite{ishwara1999} the largest source excluding 3C236, was 8C0821+695 which in our present cosmology has a projected linear size of 2.54 Mpc. In the present catalogue there are 13 sources with a size of at least 2.5 Mpc, a substantial increase that enable us to investigate the evolutionary status of these large sources. As seen in Figure \ref{rozklad} (middle panel), the FRII GRSs are mostly larger than the FRI GRSs. The median values of projected linear size of FRII, FRII/FRI and FRI type GRSs are  1.15, 1.18 and 0.99 Mpc, respectively. This indicates that the FRII radio sources tend to be larger than FRIs. Similar results were obtained by, e.g. \cite{wing2011}, who studied distributions of the projected linear size of the radio sources in clusters of galaxies. 

There has also been a substantial increase in the number of high-redshift objects, say those with z$>$1. The median redshift is 0.24 with the highest value being 3.22. There are 28 objects with a redshift of at least 1, while there were none in the earlier compilation by \cite{ishwara1999}. Although a large number of larger sources at high redshifts have been discovered from the new, more sensitive surveys, most giants are relatively nearby objects with sizes close to 1 Mpc and 1.4 GHz total luminosity logP$_{tot}$[\rm W Hz$^{-1}$] ranging from $\sim$23.0 to 28.3. The mean value of logP$_{tot}$[\rm W Hz$^{-1}$]=25.5 for giants is slightly higher than the mean 1.4 GHz total radio luminosity of smaller sized FRII's (from the sample of \citealt{koziel2011}), which is equal to logP$_{tot}$[\rm W Hz$^{-1}$]=25.1. 

\subsection{Hubble diagram for the GRSs}
\begin{figure}
\centering
    \includegraphics[width=0.99\columnwidth]{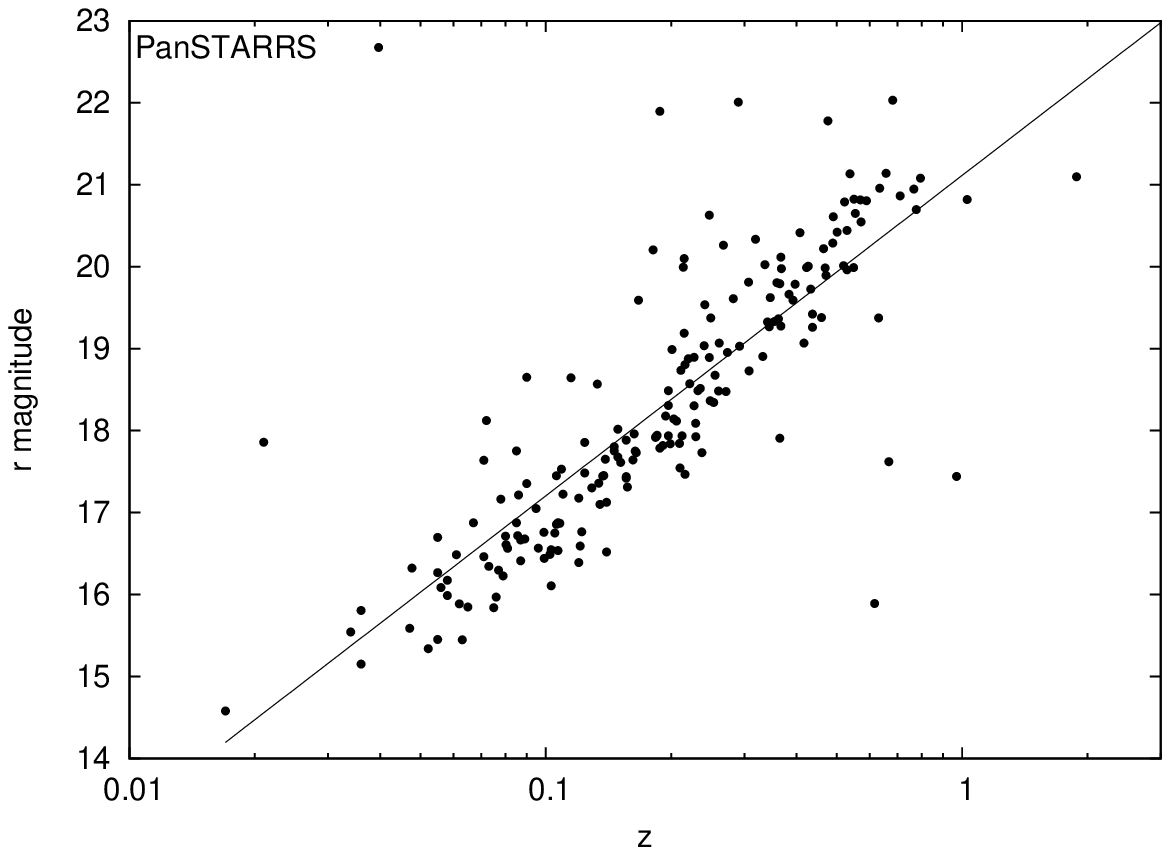}
\vspace{0.5cm}\caption{Correlation between {\it r}-band apparent PanSTARRS magnitude and redshift. The solid line corresponds to the best fit obtained for data points.}
\label{mag}
\end{figure}

Although almost all the redshifts are spectroscopic, a few listed within brackets are photometric. Figure \ref{mag} shows the {\it r}-band apparent magnitude as a function of spectroscopic redshift, for the catalogued objects. This relation is known as the ``Hubble diagram'' and can be used to estimate the ``photometric'' redshift of galaxies lacking spectroscopic observations. Subsequently, we performed a linear regression fit to the PanSTARRS data points (excluding all objects with quasar hosts from this analysis), and obtained the following relation:
\begin{equation}
r_{mag}=(3.91\pm0.18)\log(z)+(21.11\pm0.15)
\end{equation}
with a correlation coefficient of 0.89. Such a relation is consistent with the results obtained by \cite{eales1985} for a set of 3CRR galaxies, and suggests that the estimated redshifts are reasonably reliable. A similar relation but with a steeper slope, $r_{mag}=(8.83\pm0.35)$log(z)$+(22.96\pm0.37)$ was obtained for 81 large angular sized radio galaxies by \cite{lara2001b}. The differences between our and \cite{lara2001b} estimates are probably because of the different filter characteristics used during observations. For the low-redshift objects (mostly the brightest galaxies)  the magnitudes from \cite{lara2001b} are up to 3 mag brighter than those from PanSTARRS.
 
\subsection{The luminosity--linear size (P--D) diagram}
\begin{figure}
\centering
    \includegraphics[width=0.99\columnwidth]{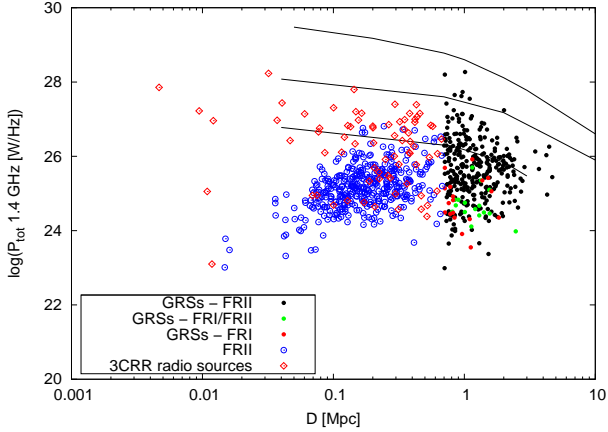}
\caption{P--D diagram for GRSs (dots), 3CRR radio sources (diamonds) from \protect\cite{laing1983}, and FRII radio sources from \protect\cite{koziel2011} (pluses). The solid lines (from top to bottom) represent the evolutionary scenarios from \protect\cite{kaiser1997} for jet powers of 1.3$\times$10$^{40}$W at z=2, 1.3$\times$10$^{39}$W at z=0.5 and 1.3$\times$10$^{38}$W at z=0.2.}
\label{pd}
\end{figure}

The radio luminosity--linear size (P--D) diagram has been used by many researchers in the past (e.g. \citealt{kaiser1997}; \citealt{ishwara1999}; \citealt{blundel1999}; \citealt{machalski2004}) as a useful tool to study the evolution of radio sources. In Figure \ref{pd} we plot the P--D relation for the sample of GRSs and smaller-sized 3CRR\footnote{http://www.jb.man.ac.uk/atlas} (\citealt{laing1983}) and FRII radio sources taken from \cite{koziel2011}. We also superimpose the evolutionary tracks proposed by \cite{kaiser1997} for three different jet powers. For the 3CRR sample the flux densities measured at 0.178 GHz were extrapolated to the frequency of 1.4 GHz using the spectral index of individual sources and then the total radio luminosities were calculated using the formula given by \cite{brown2001}. Thirteen of the radio sources from the 3CRR sample, and 21 radio sources from the FRII sample have sizes larger than 0.7 Mpc; these are included in the sample of GRSs. 

\begin{figure}
\centering
    \includegraphics[width=0.99\columnwidth]{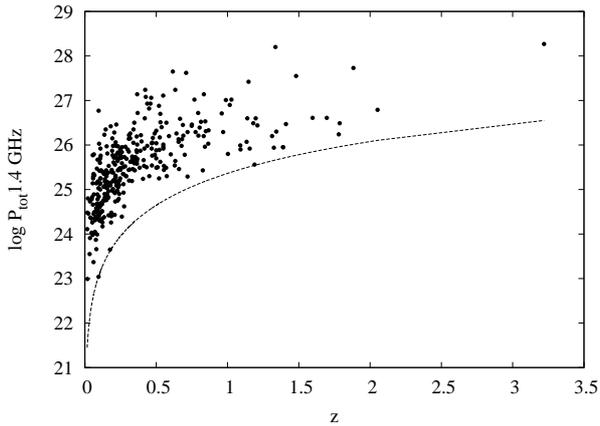}
\caption{The total radio luminosity at 1.4 GHz as a function of redshift for a sample of GRSs. The dashed line represents the total radio luminosity of the radio source with flux density of S$_{1.4 GHz}$ = 5 mJy, corresponding to the lowest flux density of sources in our catalogue.}
\label{zp}
\end{figure}

The models (e.g. \citealt{kaiser1997}; \citealt{ishwara1999}; \citealt{blundel1999}; \citealt{machalski2004}) all suggest  that the GRSs are the evolved counterparts of smaller, younger, and more luminous radio sources. The P--D diagram shows more clearly a declining upper envelope than was noted by \cite{ishwara1999}. The lack of radio sources that have both large sizes and high radio luminosities in our plot is consistent with model expectations of the evolution of radio sources. It has been postulated that GRSs evolve from normal FRII and FRI radio sources (e.g. \citealt{ishwara1999}), with sources of different jet powers following different evolutionary tracks in the P--D diagram. The upper envelope corresponds to a jet power of $\sim$10$^{40}$ W. With the discovery of many high-redshift, luminous giant sources, the  median value of total radio luminosities of the GRSs, $logP_{tot}$[\rm W Hz$^{-1}$] $\sim$ 25.5 is somewhat higher than that of typical FRII radio sources (from the sample of \citealt{koziel2011}) for which we obtained a median of $logP_{tot}$[\rm W Hz$^{-1}$]=25.1. This fact, and evolutionary tracks from \cite{kaiser1997}, indicate that the high-luminosity GRSs evolve possibly from the most luminous radio sources like those from the 3CRR sample for which the median of $logP_{tot}$[\rm W Hz$^{-1}$] is 26.4. Less-luminous GRSs can evolve from lower-luminosity FRII and FRI radio sources. 
However, we observe the deficit of low-luminosity and high-redshift GRSs visible in Figure \ref{zp} where we plot the total radio luminosity as a function of redshift, as would be expected in case of a Malmquist bias. However, in the P--D diagram (Figure \ref{pd}) just as there appears to be a rough upper envelope for the GRSs, the Figure \ref{pd} also suggests a rough lower envelope. The largest GRSs are not of the lowest luminosities amongst the GRSs. This apparent deficit of large, low-luminosity GRGs, is possibly due to a combination of selection effects (e.g. low surface-brightness of extended sources) and quenching due to inverse-Compton scattering.

\subsection{High-redshift GRSs}
\begin{figure}
\centering
    \includegraphics[width=0.99\columnwidth]{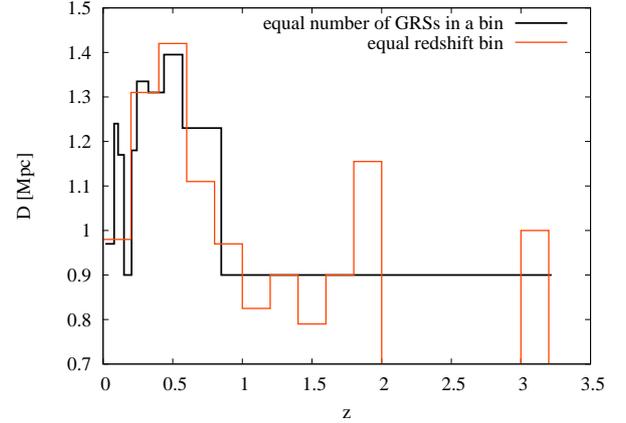}
\caption{The GRS's median size distribution as a function of redshift. Black line: the bins contain the same number of GRSs -- 35 sources each  (except the last redshift bin which has 34 sources); red line: the bins are of equal redshift -- bin size is equal to 0.2.}
\label{zd}
\end{figure}

As noted earlier, there are 28 very distant GRSs with redshift $z>1$ in our sample. However, for a long time GRSs were not expected to be found at high redshift. According to \cite{kapahi1989} since the density of the IGM increases as $\rho_{IGM}\propto(1+z)^3$, the radio lobe expansion could be significantly hampered at high redshifts. This explains the smaller median size (equal to 0.899 Mpc) of high-redshift giants in comparison with z$<$1 GRSs for which the median size is 1.16 Mpc. In addition, the surface brightness decreases with redshift as $(1+z)^{-4}$. This makes it difficult to recognise extended radio lobes and thus GRSs themselves in earlier cosmological epochs. Nevertheless, the number of high-redshift GRSs has grown due to sensitive radio surveys. 
We examine further the dependence of median size on redshift by considering equal number of GRSs in each redshift bin, and also considering redshift bins of equal width (Figure \ref{zd}). While in both cases there is a decrease in the median size with redshift for z$>$0.5, the median size increases with redshift in the range 0$<$z$<$0.5. Earlier studies of the linear size evolution of extragalactic radio sources have reported a decrease in median linear size with redshift for both bright and faint source samples (\citealt{neeser1995}; \citealt{singal1996}) of the form D$\propto(1+z)^{-n}$, where n was found to have a range of values from about 1 to 3. These were consistent with the expectations of theoretical models (\cite{kaiser1999}. However, such studies are complicated by possible correlations of size with luminosity, which may be different for galaxies and quasars (e.g. \citealt{singal1988}).  Authors (e.g. \citealt{singal1996}) have stressed the importance of examining the D--z relationship for sources of similar radio luminosity, and perhaps separately for different classes. This would be possible for a large uniformly selected sample of giant sources, which is beyond the scope of the present paper. However, it may be relevant to note the recent paper by \cite{onah2018} for a sample consisting of both galaxies and quasars, where they report an increase in median size with redshift up to about z=1. In the context of GRSs one also needs to examine whether large diffuse low-luminosity giants may be below the thresholds of existing surveys and observations.

The largest sample of GRSs within the redshift range 1$<$z$<$2 was presented by \cite{kuligowska2009}. Considering the dynamical evolution of radio galaxies, these authors concluded that the giant sizes of most distant objects are related rather to the very high power (up to $10^{40}$ W) of their jets, which overcomes the higher density of the IGM. These values are consistent with the location of these sources in the P--D diagram and the evolutionary models of radio sources.

\subsection{GRSs in regions of cosmic voids} 

Furthermore, to examine the dependence of GRSs occurrence on the IGM density, we correlated the GRS locations with the positions of 1228 cosmic voids provided by \citet{mao2017}. The catalogue of voids was compiled using the Baryon Oscillation Spectroscopic Survey (\citealt{dawson2013}), which is a part of the SDSS-III (\citealt{eisenstein2011}). The survey covers the sky area in a declination range of $-$8$\rm^{o}$$<\delta<$66$\rm^{o}$ and contains voids with Galactic latitude $|$b$|>$20$\rm^{o}$ and in a redshift range of 0.2$<$z$<$0.67. Within the sky area and redshift range covered by this void catalog, there are 91 GRGs of our compilation. However, we found just one clear correlation: the GRG J1355$+$2923 is located only about 17 Mpc from the centre (RA: $13^{h}51^{m}58\fs08$ Dec: $29\degr05\arcmin34\farcs8$, J2000.0) of a void named `CMASS North 19434' at z=0.501. The effective radius of the void is 40.5 Mpc. Although there have been a number of suggestions that GRSs occur in regions of low IGM density (e.g. \citealt{machalski2006}; \citealt{kuligowska2009}, \citealt{chen2011}, \citealt{malarecki2015}), the above result suggests that the size of GRGs is not related solely to the galaxy density around their host galaxies. Apart from deep X-ray observations, deeper optical observations including improved photometric redshifts (e.g. from PanSTARRS) could provide better estimates of the ambient galaxy density of GRGs using the methods applied by \cite{komberg2009}, but out to higher redshifts.

\section{Concluding remarks}
We present the largest-to-date sample of GRSs (defined here as having a projected largest linear size of at least 0.7 Mpc), which consists of 349 individual objects. It is limited by sensitivity and coverage of the available radio and optical surveys in different parts of the sky. Analysing the distribution of GRSs in the celestial plane, we distinguish sky regions that are underdense in GRSs and that should be explored for further objects yet to be identified even with current sensitivity limits. We estimate the lower limit to the number of GRSs on current sensitivity limits as $\sim$2000, based on the density of known GRSs in the regions where most such sources have been observed.  

The present sample has considerably increased the number of GRSs with projected sizes larger than about 2 Mpc, and also at high redshifts of greater than $\sim$1. The Hubble diagram for the galaxies in the GRS sample has been found to be similar to that of 3CRR radio galaxies, with the {\it r}-band magnitude being well correlated with redshift, allowing for quick estimates of photometric redshifts for giants that have not been spectroscopically observed.

We have examined the luminosity--linear size diagram with the enlarged sample, and find the upper envelope to be well-defined for the GRSs. This is consistent with the evolution of the radio sources with the highest jet powers of $\sim$10$^{40}$ W. The distribution is the result of different sources with different jet powers, as suggested by evolutionary scenarios (e.g. \citealt{kaiser1997}; \citealt{blundel1999};  \citealt{ishwara1999}; \citealt{machalski2004}).

The high-redshift giants that are also amongst the more luminous ones can be understood in terms of their high jet powers ($\sim$10$^{40}$ W), which overcomes the effects of higher ram pressure due to increased density of the IGM. This is consistent with the results of the P--D diagram.

While GRSs have been suggested to occur often in regions of low galaxy density, correlating our catalogue of GRSs with cosmic voids, we find only one clear association. Future X-ray telescopes will be useful for probing their environments from sensitive observations.

\section{Acknowledgments}
We thank the reviewer Prof. H. Andernach for his very detailed and valuable comments that helped improve the paper significantly. A part of this project was supported by the Polish National Center of Science under decision UMO-2016/20/S/ST9/00142.

\newpage
\begin{onecolumn}
\begin{landscape}
\begin{longtable}{l c c c c c c c c c c c c}
\caption{List of GRSs} \\                                                                                                                                                                                                                  
\hline
IAU  & $\alpha$(2000) & $\delta$(2000) & ID & z & FR & LAS & D & {\it r} & S$_{I}$ & $\delta S_{I}$ & logP$_{tot}$ & Ref. \\ 
name & (h m s) & ($\rm^{o}$ ' ") &  &  & type & (arcmin) & (Mpc) & (mag) & (mJy) & (mJy) & W$/$Hz & \\ 
(1)             & (2)                   & (3)                  & (4)                  &(5)            & (6)           & (7)           & (8)           &(9)            &(10)           & (11)          & (12)              &(13)     \\
\hline
\endfirsthead
\multicolumn{13}{l}{{\it Table 1 continued}}\\
\hline
IAU  & $\alpha$(2000) & $\delta$(2000) & ID & z & FR & LAS & D & {\it r} & S$_{I}$ & $\delta S_{I}$ & logP$_{tot}$ & Ref. \\ 
name & (h m s) & ($\rm^{o}$ ' ") &  &  & type & (arcmin) & (Mpc) & (mag) & (mJy) & (mJy) & W$/$Hz & \\ 
(1)             & (2)                   & (3)                  & (4)                  &(5)            & (6)           & (7)           & (8)           &(9)            &(10)           & (11)          & (12)              &(13)     \\
\hline
\endhead
\endfoot
\endlastfoot
\hline
J0003$+$0351	&	00 03 31.50	&	$+$03 51 11.3	&	G	&	0.0953	&	II	&	19.5	&	2.03	&	17.048	&	484.8	&	24.7	&	25.03	&	1\\
J0003$-$1517	&	00 03 49.08	&	$-$15 17 33.9	&	G	&	0.1051	&	II	&	8.3	&	0.95	&	16.751	&	480.8	&	14.4	&	25.12	&	2\\
J0010$-$1108	&	00 10 49.69	&	$-$11 08 13.0	&	G	&	0.0773	&	II	&	9.2	&	0.80	&	16.296	&	52.9	&	1.6	&	23.87	&	3\\
J0016$+$0420	&	00 16 04.36	&	$+$04 20 24.2	&	G	&	0.4328	&	II	&	6.9	&	2.40	&	19.726	&	50.0	&	5.0	&	25.50	&	4\\
J0017$-$2223	&	00 17 47.79	&	$-$22 23 19.8	&	G	&	0.1081	&	II	&	9.8	&	1.14	&	16.868	&	479.8	&	24.2	&	25.14	&	5\\
J0020$-$2016	&	00 20 33.28	&	$-$20 16 10.0	&	G	&	0.1970	&	II	&	6.2	&	1.20	&	18.488	&	907.2	&	27.3	&	25.98	&	6\\
J0022$-$0818	&	00 22 24.96	&	$-$08 18 45.7	&	G	&	0.5715	&	II	&	6.2	&	2.50	&	20.547	&	259.0	&	24.0	&	26.50	&	4,1\\
J0023$-$6710	&	00 23 40.33	&	$-$67 10 54.9	&	G	&	(1.41)	&	II	&	1.4	&	0.71	&	22.67r	&	28.1$^a$&		&	26.47	&	7\\
J0028$-$6631	&	00 28 57.83	&	$-$66 31 14.9	&	G	&	(1.78)	&	II	&	1.8	&	0.90	&	23.26r	&	9.6$^a$	&		&	26.24	&	7\\
J0031$-$6727	&	00 31 48.01	&	$-$67 27 19.9	&	G	&	1.1560	&	II	&	1.7	&	0.85	&	19.24r	&	12.7$^a$&		&	25.92	&	7\\
	&	&	&	&	&	&	&	&	&	&		&	&	\\
J0034$-$6639	&	00 34 05.63	&	$-$66 39 35.0	&	G	&	0.1075	&	II	&	16.5	&	1.93	&	16.79r	&	$101.9^b$&	6.3	&	24.47	&	8\\
J0037$+$0027	&	00 37 54.59	&	$+$00 27 17.4	&	G	&	0.5893	&	II	&	5.3	&	2.11	&	20.805	&	122.1	&	3.7	&	26.20	&	9\\
J0039$-$1300	&	00 39 39.88	&	$-$13 00 59.8	&	G	&	0.1075	&	II	&	10.7	&	1.23	&	16.856	&	153.1	&	4.6	&	24.63	&	3\\
J0041$+$3225	&	00 41 46.62	&	$+$32 25 09.6	&	G	&	(0.45)	&	II	&	2.8	&	0.97	&	19.119	&	959.4	&	28.8	&	26.82	&	10\\
J0042$-$0613	&	00 42 46.85	&	$-$06 13 52.9	&	G	&	0.1243	&	II	&	6.5	&	0.85	&	17.484	&	637.5	&	19.1	&	25.39	&	3\\
J0044$-$6656	&	00 44 46.94	&	$-$66 56 40.0	&	G	&	(0.72)	&	II	&	1.8	&	0.78	&	21.01r	&	9.3$^a$	&		&	25.29	&	7\\
J0047$+$5339	&	00 47 16.65	&	$+$53 39 36.8	&	G	&	0.1460	&	II	&	16.0	&	2.43	&	17.803	&	114.6	&	3.4	&	24.80	&	11\\
J0047$-$8308	&	00 47 55.50	&	$-$83 08 12.5	&	G	&	0.2591	&	II	&	5.6	&	1.34	&		&	$533.0^{c}$&	16.0	&	25.84	&	12\\
J0051$-$2028	&	00 51 07.10	&	$-$20 28 25.1	&	G	&	0.0856	&	II	&	8.0	&	0.76	&	16.718	&	186.6	&	9.8	&	24.52	&	5\\
J0053$+$4031	&	00 53 31.69	&	$+$40 31 25.8	&	G	&	0.1488	&	II	&	7.6	&	1.17	&	18.018	&	468.0	&	14.2	&	25.43	&	13\\
	&	&	&	&	&	&	&	&	&	&	&	&	\\
J0057$-$6606	&	00 57 26.75	&	$-$66 06 33.5	&	G	&	(1.39)	&	II	&	1.5	&	0.76	&	22.65r	&	8.9$^a$	&		&	25.95	&	7\\
J0057$+$3021	&	00 57 48.88	&	$+$30 21 08.8	&	G	&	0.0165	&	I/II	&	58.0	&	1.13	&		&	2320	&	69.8	&	24.11	&	14\\
J0104$-$6609	&	01 04 21.97	&	$-$66 09 26.7	&	G	&	(1.19)	&	II	&	1.5	&	0.75	&	22.26r	&	5.2$^a$	&		&	25.56	&	7\\
J0104$-$6704	&	01 04 31.78	&	$-$67 04 37.9	&	G	&	(1.39)	&	II	&	1.5	&	0.76	&	22.64r	&	8.8$^a$	&		&	25.95	&	7\\
J0106$-$6645	&	01 06 47.72	&	$-$66 45 51.2	&	G	&	(1.21)	&	II	&	1.8	&	0.90	&	22.29r	&	38.3$^a$&		&	26.44	&	7\\
J0107$+$3224	&	01 07 24.96	&	$+$32 24 45.2	&	G	&	0.0170	&	I	&	40.0	&	0.82	&	14.578	&	4746	&	142.4	&	24.48	&	15\\
J0109$+$7311	&	01 09 43.60	&	$+$73 11 56.0	&	G	&	0.1810	&	II	&	4.0	&	0.72	&	20.205	&	3018	&	90.5	&	26.42	&	16\\
J0112$+$4928	&	01 12 02.24	&	$+$49 28 35.0	&	G	&	0.0670	&	II	&	10.6	&	0.81	&	16.875	&	1977	&	59.4	&	25.32	&	17\\
J0115$+$2507	&	01 15 57.23	&	$+$25 07 21.0	&	G	&	0.1836	&	II	&	5.8	&	1.06	&	17.925	&	157.4	&	4.7	&	25.16	&	2\\
J0116$-$4722	&	01 16 25.05	&	$-$47 22 41.6	&	G	&	0.1461	&	II	&	11.7	&	1.78	&		&	$4077^b$&	122.9	&	26.35	&	18\\
	&	&	&	&	&	&	&	&	&	&	&	&	\\
J0117$-$0111	&	01 17 34.84	&	$-$01 11 35.6	&	Q	&	0.1855	&	II	&	5.2	&	0.96	&	19.504	&	33.0	&	2.6	&	24.48	&	1\\
J0117$+$0026	&	01 17 46.30	&	$+$00 26 22.1	&	G	&	0.5540	&	II	&	3.7	&	1.42	&	20.651	&	24.3	&	1.9	&	25.44	&	1\\
J0120$-$0038	&	01 20 12.51	&	$-$00 38 37.6	&	G	&	0.2354	&	I	&	3.2	&	0.71	&	18.515	&	310.6	&	9.4	&	25.69	&	19\\
J0129$-$0758	&	01 29 35.26	&	$-$07 58 04.3	&	G	&	0.0991	&	I/II	&	12.0	&	1.30	&	16.441	&	107.1	&	3.2	&	24.41	&	3\\
J0133$-$1303	&	01 33 13.53	&	$-$13 03 30.5	&	G	&	(0.289)	&	II	&	6.9	&	1.79	&	18.940	&	80.4	&	2.4	&	25.30	&	20\\
J0134$-$0107	&	01 34 12.78	&	$-$01 07 29.5	&	G	&	0.0790	&	I/II	&	13.7	&	1.21	&	16.226	&	258.8	&	7.8	&	24.58	&	3\\
J0135$-$0044	&	01 35 25.66	&	$-$00 44 47.3	&	G	&	0.1564	&	II	&	6.6	&	1.06	&	17.884	&	195.9	&	5.9	&	25.09	&	2\\
J0135$+$3754	&	01 35 28.33	&	$+$37 54 05.6	&	G	&	0.4373	&	II	&	2.7	&	0.91	&	19.422	&	1331	&	40.0	&	26.93	&	21\\
J0139$+$3957	&	01 39 30.46	&	$+$39 57 03.3	&	G	&	0.2107	&	II	&	5.7	&	1.17	&	18.735	&	794.5	&	23.9	&	25.99	&	22\\
J0143$-$5431	&	01 43 43.13	&	$-$54 31 39.3	&	G	&	0.1791	&	II	&	5.5	&	0.99	&		&	$217.0^{c}$&	6.5	&	25.09	&	12\\
	&	&	&	&	&	&	&	&	&	&		&	&	\\
J0152$+$0015	&	01 52 14.26	&	$+$00 15 02.6	&	G	&	0.8440	&	II	&	2.2	&	1.03	&	20.818	&	111.6	&	3.4	&	26.53	&	23\\
J0155$-$2654	&	01 55 46.30	&	$-$26 54 04.7	&	G	&	0.2091	&	II	&	3.8	&	0.77	&	17.843	&	22.3	&	0.7	&	24.43	&	24\\
J0157$+$0209	&	01 57 52.52	&	$+$02 09 54.0	&	G	&	0.2217	&	II	&	7.0	&	1.49	&	18.572	&	691.3	&	20.7	&	25.98	&	11\\
J0200$+$4048	&	02 00 30.10	&	$+$40 48 53.1	&	G	&	0.0827	&	I/II	&	14.0	&	1.30	&		&	283.5	&	9.0	&	24.67	&	13\\
J0202$-$0939	&	02 02 24.09	&	$-$09 39 00.3	&	G	&	0.7683	&	II	&	2.6	&	1.12	&	20.948	&	433.7	&	21.7	&	27.02	&	1,3\\
J0204$-$0944	&	02 04 48.29	&	$-$09 44 09.5	&	Q	&	1.0033	&	II	&	4.2	&	2.08	&	18.904	&	13.6	&	0.4	&	25.80	&	25\\
J0210$+$0118	&	02 10 08.48	&	$+$01 18 39.6	&	Q	&	0.8652	&	II	&	2.6	&	1.21	&	18.595	&	33.1	&	1.0	&	26.03	&	25\\
J0213$-$4744	&	02 13 09.53	&	$-$47 44 13.4	&	G	&	0.2200	&	II	&	6.3	&	1.33	&		&	$2115^{b}$&	64.3	&	26.46	&	18\\
J0214$+$3251	&	02 14 15.38	&	$+$32 51 05.3	&	G	&	0.2610	&	II	&	5.2	&	1.25	&	19.068	&	451.1	&	13.6	&	25.95	&	26\\
J0216$-$0449	&	02 16 59.24	&	$-$04 49 20.3	&	G	&	1.3250	&	II	&	2.4	&	1.20	&		&	9.6	&	0.3	&	25.94	&	27,28\\
	&	&	&	&	&	&	&	&	&	&	&	&	\\
J0237$-$6430	&	02 37 10.07	&	$-$64 30 01.5	&	G	&	0.3640	&	II	&	6.6	&	2.00	&		&	$154.0^{c}$&	4.6	&	25.64	&	12\\
J0241$+$0038	&	02 41 39.69	&	$+$00 38 31.7	&	G	&	0.6341	&	II	&	3.8	&	1.56	&	20.957	&	27.9	&	1.9	&	25.64	&	1\\
J0259$+$0018	&	02 59 42.88	&	$+$00 18 40.9	&	G	&	0.1834	&	II	&	4.0	&	0.73	&	17.918	&	20.3	&	2.0	&	24.26	&	29\\
J0300$-$0728	&	03 00 59.06	&	$-$07 28 30.8	&	G	&	0.4905	&	II	&	5.0	&	1.81	&	20.611	&	71.9	&	2.2	&	25.78	&	3\\
J0313$+$4120	&	03 13 01.96	&	$+$41 20 01.2	&	G	&	0.1340	&	I/II	&	9.5	&	1.34	&	17.359	&	489.8	&	14.8	&	25.35	&	30\\
J0313$-$0631	&	03 13 32.88	&	$-$06 31 57.9	&	Q	&	0.3889	&	II	&	3.5	&	1.11	&	19.325	&	151.8	&	4.6	&	25.87	&	3\\
J0315$-$0743	&	03 15 36.19	&	$-$07 43 38.8	&	G	&	(0.269)	&	II	&	3.3	&	0.86	&	19.140	&	151.0	&	16.0	&	25.50	&	4\\
J0316$-$2658	&	03 16 04.24	&	$-$26 58 06.0	&	G	&	0.2171	&	II	&	3.8	&	0.79	&	17.466	&	453.5	&	13.8	&	25.77	&	6\\
J0318$+$6829	&	03 18 18.98	&	$+$68 29 31.4	&	G	&	0.0901	&	II	&	14.9	&	1.48	&	18.650	&	777.5	&	23.5	&	25.18	&	31\\
J0320$-$4515	&	03 20 57.54	&	$-$45 15 10.7	&	G	&	0.0630	&	II	&	25.6	&	1.84	&		&	$6657^{b}$&	201.7	&	25.79	&	32\\
	&	&	&	&	&	&	&	&	&	&	&	&	\\
J0326$-$7730	&	03 26 01.33	&	$-$77 30 16.0	&	G	&	0.2771	&	II	&	5.3	&	1.33	&		&	$357.0^{c}$&	10.7	&	25.73	&	12\\
J0331$-$7713	&	03 31 39.79	&	$-$77 13 19.2	&	G	&	0.1456	&	II	&	17.7	&	2.69	&		&	$687.0^{c}$&	20.6	&	25.40	&	12\\
J0349$+$7511	&	03 49 16.28	&	$+$75 11 22.0	&	G	&	0.0803	&	II	&	14.5	&	1.30	&	16.609	&	~30	&	0.9	&	23.66	&	33\\
J0351$-$1429	&	03 51 28.54 	&	$-$14 29 08.7	&	Q 	&	0.6163 	&	II 	&	2.1	 &	0.84	& 	15.890	&	3049	&	91.5	&	27.65	&	34\\
J0401$-$8456	&	04 01 18.32	&	$-$84 56 36.1	&	G	&	0.1037	&	I	&	6.3	&	0.71	&		&	$175.0^{c}$&	5.3	&	24.49	&	12\\
J0408$-$6247	&	04 08 46.07	&	$-$62 47 51.3	&	G	&	0.0178	&	II	&	33.0	&	0.71	&		&	$210.0^{b}$&	20.0	&	22.99	&	35\\
J0422$+$1510	&	04 22 20.85	&	$+$15 10 59.5	&	G	&	0.0720	&	I	&	13.6	&	1.10	&	18.123	&	168.8	&		&	24.31	&	36\\
J0423$-$7246	&	04 23 57.57	&	$-$72 46 01.6	&	G	&	0.1921	&	II	&	4.1	&	0.78	&		&	$2143^{b}$&	65.3	&	26.15	&	18\\
J0429$+$0033	&	04 29 25.85	&	$+$00 33 04.8	&	Q	&	(0.468)	&	II	&	5.4	&	1.90	&	19.945	&	111.0	&	10.0	&	25.92	&	4\\
J0430$+$7722	&	04 30 49.49	&	$+$77 22 58.4	&	G	&	0.2150	&	II	&	4.9	&	1.02	&	19.189	&	143.8	&	4.6	&	25.27	&	37\\
	&	&	&	&	&	&	&	&	&	&	&	&	\\
J0439$-$2422	&	04 39 09.20	&	$-$24 22 08.0	&	Q	&	0.8340	&	II	&	2.1	&	0.96	&	17.782	&	454.6	&	13.7	&	27.14	&	38\\
J0443$-$6141	&	04 43 44.69	&	$-$61 41 40.2	&	Q	&	0.7200	&	II	&	1.7	&	0.73	&		&	93.3$^a$&		&	26.29	&	39\\
J0449$+$4500	&	04 49 09.08	&	$+$45 00 39.2	&	G	&	0.0208	&	I	&	30.0	&	0.76	&	17.858	&	6502	&	195.1	&	24.80	&	40\\
J0449$-$3026	&	04 49 32.35	&	$-$30 26 37.7	&	Q	&	0.3149	&	II	&	7.5	&	2.14	&	17.945	&	94.0	&	9.0	&	25.45	&	4\\
J0452$+$5204	&	04 52 52.84	&	$+$52 04 47.1	&	G	&	0.1090	&	I	&	9.7	&	1.15	&	17.530	&	2853	&	85.6	&	25.92	&	40\\
J0459$-$5250	&	04 59 16.33	&	$-$52 50 08.1	&	G	&	0.0959	&	II	&	~9	&	~0.90	&		&	$7.3^{b}$&	1.0	&	23.04	&	41\\
J0505$-$2835	&	05 05 49.22	&	$-$28 35 19.4	&	G	&	0.0381	&	II	&	40.0	&	1.79	&		&	713.9	&	21.8	&	24.37	&	42\\
J0508$+$6056	&	05 08 27.26	&	$+$60 56 27.3	&	G	&	0.0710	&	I	&	10.4	&	0.84	&	17.638	&	189.7	&	6.1	&	24.35	&	37\\
J0513$-$3028	&	05 13 31.98	&	$-$30 28 50.1	&	G	&	0.0576	&	II	&	11.4	&	0.76	&	15.987	&	2279	&	68.4	&	25.25	&	18\\
J0515$-$8059	&	05 15 55.18	&	$-$80 59 43.1	&	G	&	0.1049	&	I/II	&	7.5	&	0.86	&		&	$267.0^{c}$&	8.0	&	24.68	&	12\\
	&	&	&	&	&	&	&	&	&	&	&	&	\\
J0547$+$0108	&	05 47 13.74	&	$+$01 08 50.4	&	G	&	(0.135)$^{p}$	&	II	&	7.1	&	1.01	&	19.515	&	199.4	&	6.0	&	25.07	&	2\\
J0607$+$6114	&	06 07 34.92	&	$+$61 14 43.5	&	G	&	0.2270	&	I/II	&	5.3	&	1.15	&	18.895	&	346.4	&	10.6	&	25.70	&	37\\
J0632$-$5404	&	06 32 01.17	&	$-$54 04 57.5	&	Q	&	0.2036	&	II	&	5.2	&	1.04	&		&	$694.0^{c}$&	20.8	&	25.72	&	12\\
J0636$-$2034	&	06 36 32.26	&	$-$20 34 53.2	&	G	&	0.0552	&	II	&	15.0	&	0.95	&	16.267	&	8485	&	254.6	&	25.77	&	43\\
J0652$+$4304	&	06 52 18.38	&	$+$43 04 59.7	&	G	&	0.0891	&	II	&	9.6	&	0.95	&	16.678	&	315.7	&	9.8	&	24.78	&	26\\
J0654$+$7319	&	06 54 26.69	&	$+$73 19 50.2	&	G	&	0.1145	&	II	&	12.6	&	1.56	&	18.644	&	828.6	&	24.9	&	25.43	&	26\\
J0657$+$4808	&	06 57 51.78	&	$+$48 08 29.7	&	G	&	0.7760	&	II	&	2.2	&	0.98	&	20.697	&	78.2	&	2.6	&	26.29	&	44\\
J0702$+$4859	&	07 02 06.24	&	$+$48 59 20.1	&	G	&	0.0649	&	I/II	&	19.1	&	1.41	&	15.848	&	316.2	&	9.8	&	24.49	&	26\\
J0709$-$3601	&	07 09 14.09	&	$-$36 01 21.8	&	G	&	0.1108	&	II	&	8.2	&	0.98	&		&	1902	&	57.1	&	25.76	&	18\\
J0720$+$2837	&	07 20 14.37	&	$+$28 37 23.3	&	G	&	0.2710	&	II	&	6.3	&	1.56	&	18.477	&	39.9	&	1.8	&	24.93	&	45\\
	&	&	&	&	&	&	&	&	&	&	&	&	\\
J0725$+$3025	&	07 25 17.39	&	$+$30 25 36.3	&	G	&	(0.70)	&	II	&	2.9	&	1.24	&	19.745	&	31.0	&	1.5	&	25.78	&	46\\
J0745$+$0200	&	07 45 04.47 	&	$+$02 00 08.1 	&	G 	&	0.4650  &	II 	&	2.1	&	0.74	&	20.220	&	1535	&	46.0	&	27.06	&	47\\
J0746$-$5702	&	07 46 18.62	&	$-$57 02 58.2	&	G	&	0.1300	&	I	&	7.4	&	1.02	&		&	$194.0^{c}$&	5.8	&	24.74	&	12\\
J0748$+$5548	&	07 48 36.87	&	$+$55 48 58.3	&	G	&	0.0356	&	II	&	34.0	&	1.44	&	15.806	&	1822	&	54.7	&	24.73	&	14\\
J0750$+$6541	&	07 50 34.40	&	$+$65 41 25.6	&	Q	&	0.7490	&	II	&	3.7	&	1.63	&	17.702	&	116.7	&	3.8	&	26.43	&	37\\
J0751$+$4231	&	07 51 08.80	&	$+$42 31 24.2	&	G	&	0.2042	&	II	&	6.0	&	1.19	&	18.142	&	138.3	&	4.5	&	25.19	&	26\\
J0754$+$4316	&	07 54 07.96	&	$+$43 16 10.6	&	Q	&	0.3476	&	II	&	8.1	&	2.37	&	16.722	&	101.9	&	3.5	&	25.59	&	25,26\\
J0754$+$3033	&	07 54 48.85	&	$+$30 33 55.0	&	Q	&	0.7961	&	II	&	3.8	&	1.71	&	17.727	&	60.7	&	1.8	&	26.21	&	25\\
J0801$+$4736	&	08 01 31.97	&	$+$47 36 16.1	&	Q	&	0.1567	&	II	&	6.0	&	0.97	&	15.642	&	124.9	&	4.0	&	24.90	&	25,26\\
J0802$+$4927	&	08 02 48.80	&	$+$49 27 23.8	&	G	&	(0.678)	&	II	&	4.3	&	1.88	&	21.317	&	52.0	&	7.0	&	25.98	&	4\\
	&	&	&	&	&	&	&	&	&	&	&	&	\\
J0803$+$6656	&	08 03 45.83	&	$+$66 56 11.4	&	G	&	0.2470	&	II	&	4.7	&	1.08	&	20.629	&	174.0	&	5.5	&	25.48	&	37\\
J0807$+$7400	&	08 07 10.16	&	$+$74 00 41.8	&	G	&	0.1204	&	II	&	9.1	&	1.17	&	17.175	&	128.8	&	4.4	&	24.67	&	26\\
J0809$+$2912	&	08 09 06.22	&	$+$29 12 35.6	&	Q	&	1.4807	&	II	&	2.2	&	1.10	&	17.582	&	300.6	&	9.0	&	27.55	&	48\\
J0810$-$6800	&	08 10 55.10	&	$-$68 00 07.7	&	Q	&	0.2311	&	II	&	6.5	&	1.43	&		&	$271.0^{c}$&	8.1	&	25.43	&	12\\
J0812$+$3031	&	08 12 40.08	&	$+$30 31 09.4	&	Q	&	1.3127	&	II	&	2.4	&	1.23	&	18.533	&	18.1	&	0.5	&	26.20	&	48\\
J0813$+$4516	&	08 13 10.80	&	$+$45 16 01.4	&	G	&	0.2204	&	II	&	6.9	&	1.46	&	18.877	&	107.7	&	3.7	&	25.16	&	26\\
J0816$+$3347	&	08 16 35.52	&	$+$33 47 48.6	&	Q	&	0.5099	&	II	&	3.5	&	1.29	&	19.669	&	34.4	&	1.6	&	25.50	&	46\\
J0819$+$0549	&	08 19 41.12	&	$+$05 49 42.6	&	Q	&	1.6959	&	II	&	1.9	&	0.99	&	20.531	&	24.8	&	0.7	&	26.61	&	48\\
J0819$+$7538	&	08 19 50.50	&	$+$75 38 39.5	&	G	&	0.2324	&	II	&	8.3	&	1.83	&	18.487	&	607.8	&	18.4	&	25.97	&	26\\
J0824$+$0140	&	08 24 18.16	&	$+$01 40 40.2	&	G	&	0.2125	&	II	&	4.8	&	0.99	&	17.936	&	43.4	&	1.3	&	24.73	&	3\\
	&	&	&	&	&	&	&	&	&	&	&	&	\\
J0826$+$6920	&	08 26 01.01	&	$+$69 20 37.0	&	G	&	0.5380	&	II	&	6.7	&	2.54	&	21.133	&	238.8	&	4.6	&	26.40	&	49\\
J0842$+$2147	&	08 42 39.95	&	$+$21 47 10.3	&	Q	&	1.1814	&	II	&	2.2	&	1.08	&	18.948	&	44.9	&	1.3	&	26.49	&	48\\
J0843$-$7006	&	08 43 05.41	&	$-$70 06 56.1	&	G	&	0.1393	&	I	&	6.9	&	1.01	&	17.651	&	$203.0^{c}$&	6.1	&	24.82	&	12\\
J0843$+$2037	&	08 43 47.84	&	$+$20 37 52.4	&	Q	&	0.2276	&	II	&	5.4	&	1.17	&		&	62.0	&	1.9	&	24.96	&	50\\
J0844$+$4627	&	08 44 08.85	&	$+$46 27 44.2	&	G	&	0.5697	&	II	&	5.5	&	2.14	&	20.816	&	$82.3^{d}$&	2.5	&	25.49	&	51\\
J0856$+$6621	&	08 56 16.32	&	$+$66 21 26.8	&	G	&	0.4890	&	II	&	3.8	&	1.37	&	20.289	&	216.4	&	6.6	&	26.26	&	37\\
J0857$+$0131	&	08 57 01.76	&	$+$01 31 30.9	&	G	&	0.2734	&	II	&	5.2	&	1.30	&	18.953	&	99.0	&	9.0	&	25.34	&	4\\
J0857$+$3945	&	08 57 43.54	&	$+$39 45 28.7	&	G	&	0.5288	&	II	&	2.8	&	1.05	&	19.962	&	501.8	&	15.1	&	26.70	&	21\\
J0858$+$5620	&	08 58 32.78	&	$+$56 20 15.0	&	G	&	0.2402	&	II	&	3.9	&	0.88	&	19.035	&	77.4	&	2.3	&	25.10	&	2\\
J0902$+$5707	&	09 02 07.20	&	$+$57 07 37.9	&	Q	&	1.5964	&	II	&	1.7	&	0.87	&	18.749	&	29.3	&	0.9	&	26.61	&	52\\
	&	&	&	&	&	&	&	&	&	&	&	&	\\
J0902$+$1737	&	09 02 38.42	&	$+$17 37 51.5	&	G	&	0.1645	&	II	&	7.1	&	1.19	&	17.749	&	129.5	&	3.9	&	24.96	&	2\\
J0903$+$1208	&	09 03 03.53	&	$+$12 08 58.6	&	G	&	0.3444	&	II	&	5.3	&	1.54	&	19.267	&	107.9	&	3.2	&	25.60	&	3\\
J0908$+$3942	&	09 08 18.57	&	$+$39 42 57.1	&	G	&	1.8830	&	II	&	1.9	&	0.97	&	21.098	&	259.2	&	7.9	&	27.73	&	53\\
J0908$+$3506	&	09 08 47.89	&	$+$35 06 21.9	&	G	&	0.2600	&	II	&	6.3	&	1.51	&	18.483	&	196.6	&	6.1	&	25.59	&	26\\
J0912$+$3510	&	09 12 51.67	&	$+$35 10 12.0	&	G	&	0.2489	&	II	&	6.3	&	1.46	&	19.374	&	153.2	&	2.9	&	25.44	&	45\\
J0914$+$1006	&	09 14 19.52	&	$+$10 06 40.6	&	G	&	0.3080	&	II	&	6.3	&	1.71	&	18.729	&	483.3	&	14.5	&	26.14	&	2\\
J0918$+$2325	&	09 18 58.15	&	$+$23 25 55.4	&	Q	&	0.6899	&	II	&	2.1	&	0.89	&	17.818	&	85.9	&	2.6	&	26.21	&	25\\
J0918$+$3151	&	09 18 59.41	&	$+$31 51 40.7	&	G	&	0.0621	&	I	&	11.0	&	0.78	&	15.885	&	314.5	&	9.7	&	24.45	&	40\\
J0922$+$0919	&	09 22 21.60	&	$+$09 19 05.2	&	G	&	(0.23)	&	II	&	7.2	&	1.57	&	18.909	&	98.2	&	4.9	&	25.17	&	3\\
J0925$-$0114	&	09 25 12.73	&	$-$01 14 41.3	&	G	&	0.0730	&	II	&	13.9	&	1.14	&	16.344	&	82.5	&	2.5	&	24.02	&	3\\
	&	&	&	&	&	&	&	&	&	&	&	&	\\
J0925$+$4004	&	09 25 54.72	&	$+$40 04 14.2	&	Q	&	0.4717	&	II	&	4.4	&	1.55	&	17.891	&	75.5	&	2.3	&	25.76	&	25\\
J0926$+$6519	&	09 26 00.82	&	$+$65 19 22.7	&	G	&	0.1397	&	I	&	5.3	&	0.78	&	17.125	&	300.9	&	9.2	&	25.18	&	37\\
J0926$+$6100	&	09 26 53.38	&	$+$61 00 25.2	&	Q	&	0.2430	&	II	&	3.7	&	0.84	&	18.758	&	93.2	&	3.3	&	25.20	&	37\\
J0927$+$3511	&	09 27 49.38	&	$+$35 11 04.2	&	G	&	(0.55)	&	II	&	5.8	&	2.23	&	17.455	&	83.4	&	2.9	&	25.96	&	46\\
J0929$+$4146	&	09 29 10.67	&	$+$41 46 45.5	&	G	&	0.3650	&	II	&	6.6	&	2.00	&	19.792	&	164.6	&	5.4	&	25.84	&	26\\
J0931$+$3204	&	09 31 39.05	&	$+$32 04 00.1	&	Q	&	0.2267	&	II	&	19.9	&	4.29	&	17.487	&	65.6	&	2.0	&	24.97	&	50\\
J0932$+$1611	&	09 32 38.30	&	$+$16 11 57.2	&	G	&	0.1910	&	II	&	4.0	&	0.76	&	17.817	&	746.3	&	22.5	&	25.87	&	54\\
J0937$+$2937	&	09 37 04.04	&	$+$29 37 04.8	&	Q	&	0.4513	&	II	&	2.6	&	0.90	&	17.699	&	28.9	&	0.9	&	25.30	&	25\\
J0939$+$7405	&	09 39 47.07	&	$+$74 05 29.8	&	G	&	0.1215	&	II	&	7.3	&	0.95	&	16.765	&	87.6	&	3.3	&	24.51	&	26\\
J0939$+$3553	&	09 39 52.76	&	$+$35 53 58.9	&	G	&	0.1367	&	II	&	5.4	&	0.78	&	17.444	&	3542	&	106.3	&	26.23	&	55\\
	&	&	&	&	&	&	&	&	&	&	&	&	\\
J0944$+$2331	&	09 44 18.84	&	$+$23 31 19.8	&	Q	&	0.9890	&	II	&	1.9	&	0.91	&	17.914	&	229.1	&	6.9	&	27.01	&	25\\
J0947$-$1338	&	09 47 08.00	&	$-$13 38 27.7	&	G	&	0.0800	&	II	&	25.2	&	2.26	&	16.711	&	915.9	&	27.5	&	25.14	&	3\\
J0949$+$7314$^n$&	09 49 45.86	&	$+$73 14 23.1	&	G	&	0.0581	&	II	&	14.7	&	0.98	&	16.173	&	2522	&	75.8	&	25.29	&	37\\
J0952$+$2352	&	09 52 06.38 	&	$+$23 52 45.2 	&	Q 	&	0.9696 	&	II 	&	1.5	 &	0.70	&	17.440	&	45.1	&	1.5	&	26.29	&	25\\	
J0954$+$2715	&	09 54 19.19	&	$+$27 15 59.9	&	G	&	0.4712	&	II	&	3.9	&	1.37	&	19.895	&	146.4	&	4.4	&	26.05	&	2\\
J0959$+$1216	&	09 59 34.49	&	$+$12 16 31.5	&	Q	&	1.0895	&	II	&	2.0	&	0.98	&	18.686	&	17.1	&	0.5	&	25.99	&	25\\
J1004$+$5434	&	10 04 51.83	&	$+$54 34 04.4	&	G	&	0.0471	&	II	&	14.9	&	0.81	&	15.587	&	214.0	&	6.4	&	24.03	&	26\\
J1005$-$1315	&	10 05 57.45	&	$-$13 15 24.1	&	G	&	(0.31)	&	II	&	4.3	&	1.17	&	19.579	&	53.9	&	1.6	&	25.20	&	3\\
J1006$+$3454	&	10 06 01.74	&	$+$34 54 10.4	&	G	&	0.0994	&	II	&	39.0	&	4.23	&	16.759	&	4481	&	134.5	&	26.03	&	21\\
J1011$+$3111	&	10 11 12.11	&	$+$31 11 04.5	&	G	&	(0.50)	&	II	&	4.8	&	1.75	&	21.556	&	57.3	&	2.1	&	25.70	&	46\\
	&	&	&	&	&	&	&	&	&	&	&	&	\\
J1012$+$4229	&	10 12 44.29	&	$+$42 29 57.0	&	G	&	0.3651	&	II	&	3.1	&	0.94	&	17.906	&	89.4	&	2.7	&	25.58	&	25\\
J1014$-$0146	&	10 14 43.93	&	$-$01 46 11.9	&	G	&	0.1986	&	II	&	4.9	&	0.96	&	17.840	&	224.2	&	6.7	&	25.38	&	3\\
J1018$-$1240	&	10 18 49.83	&	$-$12 40 55.1	&	G	&	0.0793	&	I/II	&	9.3	&	0.81	&	17.163	&	226.3	&	6.8	&	24.51	&	56\\
J1020$+$0447	&	10 20 26.86	&	$+$04 47 52.0	&	Q	&	1.1338	&	II	&	1.5	&	0.74	&	19.191	&	19.0	&	0.6	&	26.07	&	25\\
J1020$+$3958	&	10 20 41.15	&	$+$39 58 11.2	&	Q	&	0.8266	&	II	&	2.7	&	1.23	&	18.092	&	9.2	&	0.3	&	25.43	&	25\\
J1021$-$0236	&	10 21 03.08	&	$-$02 36 42.6	&	G	&	0.2920	&	II	&	6.0	&	1.56	&	19.029	&	193.4	&	5.8	&	25.69	&	3\\
J1021$+$1217	&	10 21 24.22	&	$+$12 17 05.5	&	G	&	0.1294	&	II	&	14.4	&	1.97	&	17.301	&	134.5	&	4.0	&	24.75	&	3\\
J1021$+$0519	&	10 21 31.48	&	$+$05 19 00.9	&	G	&	0.1562	&	II	&	13.9	&	2.23	&	17.439	&	167.5	&	5.0	&	25.03	&	3\\
J1024$+$3647	&	10 24 43.08	&	$+$36 47 20.9	&	G	&	0.3190	&	II	&	2.9	&	0.79	&	20.335	&	132.3	&	4.0	&	25.62	&	44\\
J1027$-$2312	&	10 27 54.87	&	$-$23 12 03.4	&	Q	&	0.3090	&	II	&	3.3	&	0.89	&	18.376	&	578.5	&	17.4	&	26.22	&	38\\
	&	&	&	&	&	&	&	&	&	&	&	&	\\
J1030$+$5310	&	10 30 50.91	&	$+$53 10 28.7	&	Q	&	1.1975	&	II	&	1.7	&	0.84	&	18.013	&	55.6	&	1.7	&	26.60	&	48\\
J1032$+$2756	&	10 32 14.02	&	$+$27 56 01.7	&	G	&	0.0852	&	II	&	11.0	&	1.04	&	16.874	&	276.7	&	8.8	&	24.68	&	26\\
J1032$+$5644	&	10 32 58.90	&	$+$56 44 53.4	&	G	&	0.0453	&	I	&	35.0	&	1.83	&		&	481.8	&	14.8	&	24.35	&	40\\
J1036$+$3831	&	10 36 22.99	&	$+$38 31 30.7	&	G	&	0.3408	&	II	&	3.1	&	0.89	&	19.327	&	132.2	&	4.0	&	25.68	&	44\\
J1047$+$7419	&	10 47 54.49	&	$+$74 19 35.6	&	G	&	0.1210	&	II	&	11.2	&	1.45	&	16.592	&	59.1	&	2.9	&	24.34	&	26\\
J1048$+$1108	&	10 48 43.44	&	$+$11 08 02.4	&	G	&	0.1570	&	II	&	4.9	&	0.79	&	17.312	&	219.0	&	6.6	&	25.15	&	3\\
J1049$-$1308	&	10 49 45.73	&	$-$13 08 14.6	&	G	&	(0.15)	&	II	&	5.3	&	0.82	&	17.532	&	191.0	&	5.7	&	25.05	&	3\\
J1054$+$4152	&	10 54 03.27	&	$+$41 52 57.6	&	Q	&	1.0912	&	II	&	4.7	&	2.31	&	18.297	&	13.9	&	0.4	&	25.90	&	25\\
J1054$+$0227	&	10 54 21.16	&	$+$02 27 55.0	&	G	&	(0.34)	&	II	&	2.5	&	0.71	&	18.986	&	314.6	&	9.4	&	26.05	&	2\\
J1056$+$4100	&	10 56 36.26	&	$+$41 00 41.3	&	Q	&	1.7863	&	II	&	1.5	&	0.77	&	19.878	&	17.0	&	0.5	&	26.49	&	25\\
	&	&	&	&	&	&	&	&	&	&	&	&	\\
J1058$+$0810	&	10 58 17.69	&	$+$08 10 59.6	&	G	&	(0.33)	&	II	&	6.1	&	1.73	&	20.823	&	45.9	&	1.4	&	25.19	&	3\\
J1058$+$5140	&	10 58 17.89	&	$+$51 40 17.7	&	G	&	0.4150	&	II	&	4.5	&	1.48	&	20.035	&	59.5	&	1.8	&	25.53	&	23\\
J1058$+$2445	&	10 58 38.67	&	$+$24 45 35.1	&	G	&	(0.201)	&	II	&	10.4	&	2.13	&	18.809	&	147.0	&	12.0	&	25.21	&	4\\
J1059$-$1709	&	10 59 20.07	&	$-$17 09 22.6	&	G	&	0.1027	&	II	&	11.5	&	1.33	&	16.105	&	154	&	7.0	&	24.60	&	4\\
J1101$+$3634	&	11 01 09.46	&	$+$36 34 30.0	&	G	&	0.7500	&	II	&	2.5	&	1.10	&		&	124.0	&	3.9	&	26.46	&	44\\
J1101$-$1053	&	11 01 44.94	&	$-$10 53 18.7	&	G	&	(0.15)	&	II	&	14.2	&	2.21	&	17.948	&	98.4	&	3.0	&	24.76	&	3\\
J1101$+$4649	&	11 01 47.56	&	$+$46 49 10.4	&	G	&	0.6808	&	II	&	2.2	&	0.92	&	22.031	&	73.2	&	2.2	&	26.13	&	44\\
J1108$+$0202	&	11 08 45.49	&	$+$02 02 40.9	&	Q	&	0.1574	&	II	&	9.3	&	1.50	&	17.372	&	980.7	&	29.4	&	25.81	&	3\\
J1111$+$2657	&	11 11 25.21	&	$+$26 57 48.9	&	G	&	0.0335	&	I	&	28.7	&	1.12	&	15.543	&	135.8	&	4.1	&	23.55	&	57\\
J1113$+$4017	&	11 13 05.54	&	$+$40 17 29.8	&	G	&	0.0745	&	I/II	&	12.0	&	1.01	&	15.839	&	235.1	&	7.5	&	24.50	&	26\\
	&	&	&	&	&	&	&	&	&	&	&	&	\\
J1126$-$0042	&	11 26 03.33	&	$-$00 42 41.3	&	G	&	0.3320	&	II	&	4.0	&	1.14	&	18.905	&	160.7	&	4.8	&	25.74	&	3\\
J1130$-$1320	&	11 30 19.90	&	$-$13 20 50.0	&	Q	&	0.6337	&	II	&	4.9	&	2.01	&	16.166	&	1134	&	34.1	&	27.24	&	58\\
J1130$+$0629	&	11 30 51.73	&	$+$06 29 53.3	&	G	&	0.3970	&	II	&	4.4	&	1.41	&	19.785	&	232.3	&	7.0	&	26.08	&	1,2\\
J1145$-$0033	&	11 45 53.66	&	$-$00 33 04.5	&	Q	&	2.0522	&	II	&	2.6	&	1.34	&	19.628	&	24.2	&	0.7	&	26.79	&	59\\
J1147$+$3501	&	11 47 22.13	&	$+$35 01 07.5	&	G	&	0.0631	&	II	&	11.8	&	0.85	&	15.447	&	842.0	&	25.4	&	24.89	&	26\\
J1148$-$0404	&	11 48 55.88	&	$-$04 04 09.6	&	Q	&	0.3410	&	II	&	3.2	&	0.93	&	18.371	&	612.9	&	18.5	&	26.35	&	60\\
J1151$+$3355	&	11 51 39.68	&	$+$33 55 41.8	&	Q	&	0.8481	&	II	&	2.1	&	0.97	&	18.207	&	65.2	&	2.0	&	26.31	&	25\\
J1155$+$4029	&	11 55 49.54	&	$+$40 29 40.2	&	G	&	(0.53)	&	II	&	3.8	&	1.43	&	21.635	&	317.6	&	9.6	&	26.51	&	46\\
J1200$+$3449	&	12 00 50.48	&	$+$34 49 20.9	&	G	&	(0.50)	&	II	&	2.5	&	0.91	&	21.517	&	234.0	&	7.2	&	26.32	&	46\\
J1208$+$0259	&	12 08 17.53	&	$+$02 59 16.5	&	G	&	0.2410	&	II	&	3.5	&	0.73	&	19.536	&	47.9	&	2.8	&	24.90	&	1\\
	&	&	&	&	&	&	&	&	&	&	&	&	\\
J1211$+$7419	&	12 11 58.68	&	$+$74 19 04.1	&	G	&	0.1076	&	II	&	8.1	&	0.94	&	16.535	&	625.1	&	18.9	&	25.25	&	61\\
J1213$-$0500	&	12 13 32.87	&	$-$05 00 24.1	&	G	&	0.0860	&	II	&	8.9	&	0.85	&	17.215	&	55.7	&	1.7	&	23.99	&	3\\
J1216$+$4159	&	12 16 09.60	&	$+$41 59 28.3	&	Q	&	0.2426	&	II	&	5.2	&	1.19	&	18.350	&	401.5	&	12.5	&	25.83	&	26\\
J1216$+$6724	&	12 16 37.24	&	$+$67 24 41.4	&	G	&	0.3616	&	II	&	5.8	&	1.75	&	19.365	&	181.2	&	5.7	&	25.88	&	37\\
J1220$+$6341	&	12 20 36.41	&	$+$63 41 44.4	&	G	&	0.1876	&	II	&	4.7	&	0.88	&	17.785	&	255.0	&	7.9	&	25.39	&	62\\
J1229$+$3555	&	12 29 25.53	&	$+$35 55 32.1	&	Q	&	0.8328	&	II	&	1.7	&	0.78	&	19.375	&	54.2	&	1.6	&	26.20	&	25\\
J1232$-$1015	&	12 32 15.90	&	$-$10 15 25.6	&	G	&	0.2531	&	II	&	8.6	&	1.98	&	18.344	&	302.6	&	15.4	&	25.75	&	1\\
J1234$+$5318	&	12 34 58.46	&	$+$53 18 51.3	&	G	&	(0.642)	&	II	&	10.8	&	4.45	&	20.690	&	114.8	&	6.1	&	26.26	&	26\\
J1235+3925$^m$	&	12 35 04.75	&	$+$39 25 38.1	&	G	&	3.2200	&	II	&	2.2	&	1.01	&		&	260.1	&	8.0	&	28.27	&	63\\
J1235$+$2120	&	12 35 26.67	&	$+$21 20 34.8	&	G	&	0.4227	&	II	&	2.5	&	0.83	&	19.990	&	2911	&	87.4	&	27.24	&	21\\
	&	&	&	&	&	&	&	&	&	&	&	&	\\
J1236$+$1034	&	12 36 04.52 	&	$+$10 34 49.3	&	Q 	&	0.6671 	&	II 	&	1.7	 &	0.71	&	17.620	&	222.7	&	6.7	&	26.59	&	25\\
J1242$+$3838	&	12 42 36.82	&	$+$38 38 06.1	&	G	&	0.4077	&	II	&	2.3	&	0.74	&	20.415	&	31.9	&	1.0	&	25.24	&	64\\
J1247$+$6723	&	12 47 33.33	&	$+$67 23 16.5	&	G	&	0.1073	&	II	&	11.6	&	1.35	&	16.876	&	356.7	&	10.9	&	25.00	&	26\\
J1251$+$7537	&	12 51 05.99	&	$+$75 37 38.9	&	G	&	0.1970	&	II	&	4.0	&	0.78	&	17.936	&	86.7	&	3.0	&	24.96	&	37\\
J1251$+$5034	&	12 51 42.03 	&	$+$50 34 24.7	&	G 	&	0.5490 	&	II 	&	2.3	 &	0.88	&	19.990	&	1175	&	35.2	&	27.11	&	65\\
J1253$-$0139	&	12 53 09.09	&	$-$01 39 39.9	&	G	&	(0.24)	&	II	&	3.9	&	0.88	&		&	176.2	&	5.3	&	25.46	&	3\\
J1253$+$4041	&	12 53 12.28	&	$+$40 41 24.6	&	G	&	0.2295	&	I/II	&	4.6	&	1.01	&	18.090	&	38.2	&	1.8	&	24.75	&	45\\
J1254$+$2933	&	12 54 34.05	&	$+$29 33 41.0	&	G	&	0.5290	&	II	&	4.9	&	1.84	&	20.443	&	41.4	&	1.9	&	25.62	&	46\\
J1256$+$3638	&	12 56 13.90	&	$+$36 38 29.1	&	G	&	0.4685	&	II	&	2.6	&	0.90	&	19.986	&	145.3	&	4.4	&	26.04	&	44\\
J1259$-$7737	&	12 59 09.00	&	$-$77 37 28.4	&	G	&	($>$0.3)&	II	&	5.7	&	$>$1.51	&		&	$439.0^{c}$&	13.2	&	25.90	&	12\\
	&	&	&	&	&	&	&	&	&	&	&	&	\\
J1301$+$5105	&	13 01 25.90 	&	$+$51 05 00.7   &	G 	&	(0.275) &	II 	&	10.3	 &	2.57	&	19.050	&	18.7	&	2.8	&	24.62	&	23\\
J1304$+$2454	&	13 04 51.42	&	$+$24 54 45.9	&	Q	&	0.6025	&	II	&	2.4	&	0.97	&	17.497	&	45.0	&	1.4	&	25.79	&	25\\
J1304$-$3249	&	13 04 58.53	&	$-$32 49 15.9	&	G	&	0.1530	&	II	&	6.0	&	0.95	&		&	1530	&	46.0	&	25.97	&	21\\
J1308$+$6154	&	13 08 44.75	&	$+$61 54 15.3	&	G	&	0.1626	&	II	&	8.9	&	1.48	&	17.642	&	94.9	&	3.5	&	24.81	&	26\\
J1311$-$4422	&	13 11 23.76	&	$-$44 22 41.2	&	G	&	0.0506	&	I/II	&	14.4	&	0.85	&		&	$1119^{b}$&	35.5	&	24.83	&	66\\
J1311$+$4058	&	13 11 43.09	&	$+$40 58 59.8	&	G	&	0.1106	&	II	&	6.2	&	0.74	&	17.225	&	594.3	&	18.0	&	25.25	&	67\\
J1312$+$4450	&	13 12 17.00	&	$+$44 50 21.3	&	G	&	0.0356	&	I	&	22.6	&	0.96	&	15.151	&	278.4	&	8.6	&	23.91	&	26\\
J1313$+$6937	&	13 13 58.86	&	$+$69 37 18.2	&	G	&	0.1060	&	II	&	7.1	&	0.82	&	17.449	&	1433	&	43.1	&	25.60	&	62\\
J1321$+$3741	&	13 21 06.64	&	$+$37 41 53.5	&	Q	&	1.1390	&	II	&	1.5	&	0.74	&	18.776	&	63.8	&	1.9	&	26.60	&	25\\
J1326$+$4934	&	13 26 14.34	&	$+$49 34 31.5	&	G	&	(0.42)	&	II	&	2.8	&	0.93	&	19.615	&	674.9	&	20.3	&	26.60	&	21\\
	&	&	&	&	&	&	&	&	&	&	&	&	\\
J1327$+$5749	&	13 27 41.32	&	$+$57 49 43.4	&	G	&	0.1202	&	II	&	12.1	&	1.61	&	16.390	&	89.0	&	4.0	&	24.51	&	4\\
J1327$+$1748	&	13 27 43.50	&	$+$17 48 37.5	&	G	&	0.6569	&	II	&	2.0	&	0.84	&	21.139	&	70.0	&	12.0	&	26.07	&	4\\
J1328$-$0129	&	13 28 34.15	&	$-$01 29 17.6	&	G	&	0.1513	&	II	&	5.4	&	0.85	&	17.613	&	335.2	&	10.1	&	25.30	&	3\\
J1328$-$0307	&	13 28 34.37	&	$-$03 07 44.8	&	G	&	0.0854	&	II	&	13.4	&	1.28	&	17.752	&	214.2	&	7.0	&	24.57	&	45\\
J1330$+$1826	&	13 30 12.44	&	$+$18 26 06.8	&	Q	&	0.2592	&	II	&	3.5	&	0.84	&	18.368	&	12.5	&	0.5	&	24.39	&	50\\
J1330$+$3850	&	13 30 36.19	&	$+$38 50 19.7	&	G	&	0.6300	&	II	&	4.7	&	1.93	&	19.375	&	22.5	&	1.4	&	25.54	&	46\\
J1334$-$1009	&	13 34 18.56	&	$-$10 09 29.0	&	G	&	0.0838	&	II	&	13.7	&	1.24	&	16.564	&	1878	&	56.4	&	25.47	&	68\\
J1335$-$8018	&	13 35 59.70	&	$-$80 18 05.1	&	G	&	0.2478	&	II	&	10.1	&	2.34	&		&	$483.0^{c}$&	14.5	&	25.75	&	12\\
J1340$+$4232	&	13 40 34.70	&	$+$42 32 32.2	&	Q	&	1.3424	&	II	&	2.3	&	1.17	&	18.929	&	21.6	&	0.7	&	26.30	&	25\\
J1342$+$3758	&	13 42 54.54	&	$+$37 58 18.1	&	G	&	0.2270	&	II	&	11.3	&	2.45	&	18.304	&	104.9	&	3.5	&	25.18	&	45\\
	&	&	&	&	&	&	&	&	&	&	&	&	\\
J1344$+$3317	&	13 44 15.75	&	$+$33 17 19.1	&	Q	&	0.6868	&	II	&	2.4	&	1.04	&	19.186	&	149.5	&	4.5	&	26.45	&	2\\
J1345$+$5403	&	13 45 57.56	&	$+$54 03 16.6	&	G	&	0.1626	&	II	&	4.8	&	0.80	&	17.960	&	340.3	&	10.5	&	25.38	&	21\\
J1350$-$1634	&	13 50 36.14	&	$-$16 34 49.5	&	Q	&	0.0977	&	II	&	11.1	&	1.19	&	16.849	&	299.6	&	9.0	&	24.84	&	2\\
J1350$+$6429	&	13 50 42.04	&	$+$64 29 30.6	&	G	&	0.7100	&	II	&	2.2	&	0.95	&	20.864	&	2036	&	61.1	&	27.62	&	69\\
J1353$+$2631	&	13 53 35.92	&	$+$26 31 47.5	&	Q	&	0.3079	&	II	&	2.6	&	0.71	&	17.488	&	244.6	&	7.3	&	25.85	&	21\\
J1354$-$0705	&	13 54 18.21	&	$-$07 05 27.1	&	G	&	(0.19)	&	II	&	4.0	&	0.75	&	18.947	&	97.9	&	2.9	&	24.98	&	3\\
J1355$+$2923	&	13 55 17.61	&	$+$29 23 33.9	&	G	&	0.5010	&	II	&	4.4	&	1.61	&	20.422	&	123.0	&	3.9	&	26.04	&	46\\
J1400$+$3019	&	14 00 43.43	&	$+$30 19 18.7	&	G	&	0.2060	&	II	&	10.9	&	2.19	&	18.117	&	446.8	&	13.7	&	25.72	&	70\\
J1408$+$3054	&	14 08 06.21	&	$+$30 54 48.4	&	Q	&	0.8413	&	II	&	3.7	&	1.69	&	17.322	&	30.1	&	0.9	&	25.96	&	71\\
J1409$-$0302	&	14 09 48.86	&	$-$03 02 32.6	&	G	&	0.1378	&	II	&	9.5	&	1.37	&	17.452	&	164.0	&	4.9	&	24.90	&	72\\
	&	&	&	&	&	&	&	&	&	&	&	&	\\
J1410$+$2955	&	14 10 36.80	&	$+$29 55 50.9	&	Q	&	0.5740	&	II	&	2.5	&	0.98	&	18.540	&	16.4	&	0.5	&	25.30	&	25\\
J1411$+$0619	&	14 11 09.74	&	$+$06 19 45.6	&	G	&	0.3590	&	II	&	5.2	&	1.56	&	19.805	&	110.5	&	3.3	&	25.65	&	3\\
J1412$+$4212	&	14 12 44.10	&	$+$42 12 57.7	&	Q	&	0.8046	&	II	&	1.7	&	0.78	&	19.887	&	89.9	&	2.7	&	26.39	&	44\\
J1418$+$3746	&	14 18 37.65	&	$+$37 46 24.5	&	G	&	0.1349	&	II	&	7.7	&	1.09	&	17.099	&	51.8	&	3.4	&	24.38	&	26\\
J1420$-$0545	&	14 20 23.80	&	$-$05 45 28.8	&	G	&	0.3067	&	II	&	17.4	&	4.69	&	19.811	&	95.5	&	2.9	&	25.44	&	73\\
J1421$+$4144	&	14 21 05.61	&	$+$41 44 48.5	&	G	&	0.3670	&	II	&	3.2	&	0.97	&	19.275	&	3177	&	95.3	&	27.14	&	63\\
J1427$+$2632	&	14 27 35.61	&	$+$26 32 14.5	&	Q	&	0.3638	&	II	&	4.0	&	1.21	&	16.409	&	367.3	&	11.2	&	26.19	&	21\\
J1428$+$2918	&	14 28 19.24	&	$+$29 18 44.2	&	G	&	0.0870	&	II	&	14.7	&	1.42	&	16.411	&	430.7	&	13.4	&	24.89	&	26\\
J1428$+$3938	&	14 28 45.98	&	$+$39 38 42.2	&	G	&	(0.50)	&	II	&	4.5	&	1.64	&	20.812	&	86.8	&	3.1	&	25.88	&	46\\
J1429$+$0715	&	14 29 55.38	&	$+$07 15 12.9	&	G	&	0.0548	&	I	&	11.2	&	0.71	&	15.452	&	1969	&	59.2	&	25.14	&	74\\
	&	&	&	&	&	&	&	&	&	&	&	&	\\
J1432$+$1548	&	14 32 15.54	&	$+$15 48 22.4	&	Q	&	1.0159	&	II	&	2.8	&	1.36	&	18.456	&	165.9	&	5.2	&	26.90	&	60,75\\
J1435$+$4948	&	14 35 10.30	&	$+$49 48 19.4	&	Q	&	0.1661	&	II	&	5.0	&	0.84	&	17.461	&	36.0	&	1.1	&	24.42	&	76\\
J1436$-$1613	&	14 36 49.61	&	$-$16 13 41.0	&	Q	&	0.1445	&	I/II	&	10.3	&	1.55	&	15.803	&	235.2	&	9.8	&	25.10	&	77\\
J1445$+$3051	&	14 45 27.05	&	$+$30 51 28.9	&	G	&	0.4169	&	II	&	5.0	&	1.65	&	19.068	&	142.9	&	4.7	&	25.92	&	46\\
J1445$-$0540	&	14 45 56.35	&	$-$05 40 56.7	&	G	&	0.3670	&	II	&	7.0	&	2.13	&	20.117	&	239.6	&	7.2	&	26.01	&	3\\
J1448$-$4008	&	14 48 51.01	&	$-$40 08 45.7	&	G	&	0.1230	&	II	&	11.5	&	1.54	&		&	48.0	&		&	24.26	&	78,79\\
J1450$+$1006	&	14 50 49.40	&	$+$10 06 49.1	&	G	&	0.0545	&	I/II	&	39.0	&	2.47	&	16.698	&	137.5	&	5.2	&	23.98	&	80\\
J1451$+$3357	&	14 51 33.58	&	$+$33 57 42.8	&	G	&	0.3251	&	II	&	4.1	&	1.15	&		&	131.3	&	4.3	&	25.63	&	46\\
J1453$+$3308	&	14 53 02.86	&	$+$33 08 42.4	&	G	&	0.2482	&	II	&	5.6	&	1.30	&	18.365	&	450.7	&	13.5	&	25.90	&	26,64,81\\
J1457$-$0613	&	14 57 47.12	&	$-$06 13 18.0	&	G	&	0.1671	&	II	&	4.2	&	0.71	&	19.591	&	465.3	&	14.0	&	25.53	&	3\\
	&	&	&	&	&	&	&	&	&	&	&	&	\\
J1459$-$0432	&	14 59 57.24	&	$-$04 32 29.8	&	G	&	(0.27)	&	II	&	6.8	&	1.68	&	19.239	&	61.3	&	1.8	&	25.12	&	3\\
J1504$+$6856	&	15 04 12.77	&	$+$68 56 12.8	&	Q	&	0.3180	&	II	&	3.1	&	0.87	&	17.578	&	458.2	&	13.7	&	26.15	&	2,37\\
J1507$+$0234	&	15 07 03.78	&	$+$02 34 07.2	&	G	&	0.1238	&	II	&	6.3	&	0.83	&	17.856	&	38.0	&	2.5	&	24.17	&	29\\
J1511$+$0751	&	15 11 00.01 	&	$+$07 51 50.0 	&	G 	&	0.4594 	&	II 	&	2.2	 &	0.76	&	19.380	&	1175	&	35.2	&	26.93	&	82\\
J1513$+$3841	&	15 13 29.65	&	$+$38 41 57.3	&	G	&	(0.52)	&	II	&	4.5	&	1.68	&	19.041	&	18.2	&	1.3	&	25.25	&	48\\
J1520$-$0546	&	15 20 13.29	&	$-$05 46 27.0	&	G	&	0.0605	&	II	&	22.0	&	1.53	&	16.485	&	26.8	&	0.8	&	23.37	&	3\\
J1521$+$5105	&	15 21 14.55	&	$+$51 05 00.9	&	G	&	(0.37)	&	II	&	4.3	&	1.31	&	19.139	&	1198	&	36.0	&	26.72	&	65\\
J1525$+$3345	&	15 25 00.79	&	$+$33 45 42.4	&	G	&	(0.47)	&	II	&	3.6	&	1.27	&	20.974	&	44.9	&	2.0	&	25.54	&	46\\
J1536$+$8423	&	15 36 57.27	&	$+$84 23 10.8	&	G	&	0.2010	&	II	&	7.9	&	1.56	&	18.989	&	368.0	&	11.3	&	25.61	&	26\\
J1540$-$0127	&	15 40 56.80	&	$-$01 27 10.0	&	G	&	0.1490	&	II	&	4.9	&	0.76	&	17.679	&	194.1	&	5.8	&	25.05	&	3\\
	&	&	&	&	&	&	&	&	&	&	&	&	\\
J1543$-$0112	&	15 43 11.95	&	$-$01 12 46.6	&	G	&	0.3680	&	II	&	2.9	&	0.88	&	19.977	&	114.3	&	3.4	&	25.69	&	3\\
J1548$-$3216	&	15 48 58.05	&	$-$32 16 57.6	&	G	&	0.1082	&	II	&	8.3	&	0.97	&		&	1772	&	53.3	&	25.71	&	18\\
J1552$+$2005	&	15 52 09.19	&	$+$20 05 23.2	&	G	&	0.0890	&	II	&	19.6	&	1.95	&	17.353	&	2384	&	71.6	&	25.67	&	14\\
J1554$+$3945	&	15 54 26.85	&	$+$39 45 08.7	&	G	&	(0.35)	&	II	&	3.7	&	1.09	&	19.997	&	69.3	&	2.5	&	25.43	&	46\\
J1555$+$3653	&	15 55 00.39	&	$+$36 53 37.4	&	G	&	0.2472	&	II	&	5.8	&	1.34	&	18.891	&	100.9	&	3.5	&	25.25	&	46\\
J1604$+$3731	&	16 04 23.44	&	$+$37 31 49.3	&	G	&	0.8140	&	II	&	2.9	&	1.32	&		&	117.0	&	3.8	&	26.52	&	44\\
J1604$+$3438	&	16 04 45.89	&	$+$34 38 16.5	&	G	&	0.2817	&	II	&	3.3	&	0.84	&	19.611	&	139.6	&	4.5	&	25.52	&	46\\
J1615$+$3826	&	16 15 52.23	&	$+$38 26 31.9	&	G	&	0.1856	&	II	&	4.4	&	0.81	&	17.942	&	27.0	&	1.5	&	24.40	&	46\\
J1616$+$4825	&	16 16 01.26	&	$+$48 25 35.4	&	G	&	(0.233)	&	II	&	12.3	&	2.71	&	18.954	&	73.5	&	4.3	&	25.05	&	26\\
J1628$+$5146	&	16 28 04.06	&	$+$51 46 31.4	&	G	&	0.0560	&	II	&	18.4	&	1.19	&	16.085	&	628.8	&	19.1	&	24.66	&	83\\
	&	&	&	&	&	&	&	&	&	&	&	&	\\
J1632$+$8232	&	16 32 31.97	&	$+$82 32 16.4	&	G	&	0.0247	&	I/II	&	52.0	&	1.56	&		&	2100	&	63.1	&	24.46	&	18\\
J1635$+$3608	&	16 35 22.54	&	$+$36 08 04.9	&	G	&	0.1650	&	I/II	&	5.3	&	0.90	&	17.731	&	95.1	&	3.3	&	24.83	&	46\\
J1637$+$4146	&	16 37 53.38	&	$+$41 46 01.4	&	G	&	0.8670	&	II	&	2.2	&	1.02	&		&	64.7	&	2.3	&	26.33	&	44\\
J1649$+$3114	&	16 49 06.12	&	$+$31 14 31.4	&	G	&	0.4373	&	II	&	3.5	&	1.18	&	19.261	&	146.8	&	4.7	&	25.98	&	46\\
J1702$+$4217	&	17 02 55.95	&	$+$42 17 48.8	&	G	&	0.4760	&	II	&	3.0	&	1.07	&	21.780	&	181.6	&	5.7	&	26.16	&	44\\
J1706$+$4340	&	17 06 25.44	&	$+$43 40 40.2	&	Q/G	&	(0.525)	&	II	&	2.0	&	0.75	&	19.743	&	148.0	&	4.4	&	26.17	&	84\\
J1712$+$3558	&	17 12 24.87	&	$+$35 58 26.2	&	G	&	0.3357	&	II	&	3.5	&	1.00	&	20.025	&	82.6	&	2.9	&	25.46	&	46\\
J1723$+$3417	&	17 23 20.80	&	$+$34 17 58.0	&	Q	&	0.2060	&	II	&	4.1	&	0.82	&	16.043	&	1639	&	49.2	&	26.28	&	38\\
J1725$+$3923	&	17 25 17.05	&	$+$39 23 05.3	&	G	&	0.2898	&	II	&	4.8	&	1.24	&	22.008	&	78.8	&	3.0	&	25.30	&	46\\
J1728$-$7237	&	17 28 28.10	&	$-$72 37 34.9	&	G	&	0.4735	&	II	&	6.2	&	2.20	&		&	$214.0^{c}$&	6.4	&	26.05	&	12\\
	&	&	&	&	&	&	&	&	&	&	&	&	\\
J1738$+$3733	&	17 38 20.93	&	$+$37 33 33.8	&	G	&	0.1562	&	II	&	6.5	&	1.04	&	17.419	&	235.5	&	7.3	&	25.17	&	26\\
J1745$+$7115	&	17 45 43.53	&	$+$71 15 48.7	&	G	&	0.2160	&	II	&	4.4	&	0.92	&	18.805	&	887.5	&	26.7	&	26.06	&	37\\
J1748$-$2335	&	17 48 39.05	&	$-$23 35 21.2	&	G	&	0.2400	&	II	&	5.7	&	1.29	&		&	398.5	&	12.0	&	25.81	&	78,85\\
J1815$+$6531	&	18 15 11.79	&	$+$65 31 21.9	&	G	&	0.9600	&	II	&	2.1	&	1.00	&		&	149.8	&	4.7	&	26.71	&	86,87\\
J1815$+$6818	&	18 15 29.52	&	$+$68 18 30.5	&	G	&	0.7940	&	II	&	3.3	&	1.49	&	21.081	&	195.8	&	6.1	&	26.72	&	86,87\\
J1835$+$6635	&	18 35 07.30	&	$+$66 35 00.0	&	G	&	0.3540	&	II	&	4.4	&	1.31	&	19.331	&	136.7	&	4.5	&	25.73	&	37\\
J1835$+$6204	&	18 35 10.92	&	$+$62 04 08.1	&	G	&	0.5194	&	II	&	3.8	&	1.42	&	20.013	&	794.5	&	23.9	&	26.88	&	88\\
J1838$+$6555	&	18 38 11.54	&	$+$65 55 02.6	&	Q	&	0.2300	&	II	&	6.8	&	1.49	&	16.638	&	454.6	&	22.8	&	25.83	&	26,89\\
J1844$+$6522	&	18 44 07.41	&	$+$65 22 03.1	&	G	&	0.1970	&	II	&	7.5	&	1.45	&	18.308	&	98.0	&	3.6	&	25.01	&	37\\
J1853$+$5046	&	18 53 32.51	&	$+$50 46 07.3	&	G	&	0.0958	&	II	&	7.0	&	0.74	&	16.566	&	126.5	&	4.4	&	24.45	&	26\\
	&	&	&	&	&	&	&	&	&	&	&	&	\\
J1853$+$8002	&	18 53 51.07	&	$+$80 02 42.5	&	G	&	0.2139	&	II	&	5.6	&	1.16	&	19.994	&	153.9	&	5.0	&	25.29	&	37\\
J1910$-$7049	&	19 10 55.20	&	$-$70 49 00.9	&	G	&	0.2152	&	II	&	6.0	&	1.25	&		&	$238.0^{c} $&	7.1	&	25.31	&	12\\
J1918$+$7415	&	19 18 34.89	&	$+$74 15 05.1	&	G	&	0.1940	&	II	&	6.6	&	1.26	&	18.179	&	571.7	&	17.3	&	25.77	&	37\\
J1919$-$7959	&	19 19 13.86	&	$-$79 59 06.7	&	G	&	0.3460	&	II	&	6.1	&	1.78	&	19.622	&	$1262^{c}$&	37.9	&	26.50	&	18\\
J1919$+$5143	&	19 19 22.75	&	$+$51 43 33.9	&	Q	&	0.2840	&	II	&	7.3	&	1.86	&		&	385.3	&	11.8	&	25.96	&	26\\
J1920$+$4526	&	19 20 01.66	&	$+$45 26 52.9	&	G	&	0.0522	&	II	&	16.7	&	1.01	&	15.340	&	287.9	&	14.9	&	24.26	&	26\\
J1921$+$4806	&	19 21 13.97	&	$+$48 06 18.7	&	G	&	0.1023	&	I	&	13.5	&	1.52	&	16.489	&	1032	&	51.7	&	25.42	&	26\\
J1946$-$8222	&	19 46 50.50	&	$-$82 22 53.8	&	G	&	0.3330	&	II	&	7.4	&	2.11	&		&	$219.0^{c}$&	6.6	&	25.70	&	12\\
J1951$+$7037	&	19 51 40.82	&	$+$70 37 40.0	&	G	&	0.5500	&	II	&	5.2	&	2.00	&	20.825	&	92.3	&	3.2	&	26.01	&	37\\
J2008$+$0049	&	20 08 43.37	&	$+$00 49 18.9	&	G	&	(0.412)	&	II	&	11.0	&	3.71	&	20.526	&	81.0	&	3.0	&	25.66	&	4\\
	&	&	&	&	&	&	&	&	&	&	&	&	\\
J2018$-$5539	&	20 18 01.31	&	$-$55 39 30.8	&	G	&	0.0606	&	I	&	~20	&	1.37	&		&	$2653^{b}$&	80.9	&	25.35	&	66,90\\
J2034$-$2630	&	20 34 49.23	&	$-$26 30 36.4	&	G	&	0.1033	&	II	&	6.7	&	0.78	&	16.547	&	76.0	&	6.0	&	24.30	&	4\\
J2035$+$6805	&	20 35 16.55	&	$+$68 05 41.6	&	G	&	0.1330	&	I	&	11.5	&	1.61	&	18.568	&	253.7	&	8.1	&	25.06	&	37\\
J2042$+$7508	&	20 42 37.31	&	$+$75 08 02.4	&	Q	&	0.1040	&	II	&	10.2	&	1.16	&	14.576	&	1789	&	53.7	&	25.68	&	91\\
J2059$+$6247	&	20 59 09.56	&	$+$62 47 44.1	&	G	&	0.2670	&	II	&	4.7	&	1.15	&	20.262	&	107.7	&	3.6	&	25.35	&	37\\
J2059$+$2434	&	20 59 39.81	&	$+$24 34 23.9	&	G	&	(0.116)	&	II	&	8.2	&	1.06	&	17.740	&	149.0	&	8.0	&	24.70	&	4\\
J2103$+$6456	&	21 03 13.87	&	$+$64 56 55.3	&	G	&	0.2150	&	II	&	4.8	&	1.00	&	20.100	&	118.3	&	4.0	&	25.18	&	37\\
J2145$+$8154	&	21 45 30.90	&	$+$81 54 53.7	&	G	&	0.1457	&	II	&	18.3	&	2.78	&	17.753	&	395.4	&	12.2	&	25.34	&	26\\
J2159$-$7219	&	21 59 09.99	&	$-$72 19 01.1	&	G	&	0.0970	&	II	&	9.2	&	0.98	&		&	$142.0^{c}$&	4.3	&	24.34	&	12\\
J2219$-$2021	&	22 19 44.23	&	$-$20 21 30.6	&	G	&	1.1480	&	II	&	1.5	&	0.73	&		&	407.6	&	12.2	&	27.42	&	6\\
	&	&	&	&	&	&	&	&	&	&	&	&	\\
J2230$-$3942	&	22 30 40.28	&	$-$39 42 52.1	&	Q	&	0.3181	&	II	&	3.3	&	0.91	&		&	591.1	&	17.7	&	26.26	&	2\\
J2233$+$1315	&	22 33 01.30	&	$+$13 15 02.5	&	G	&	(0.093)	&	II	&	16.0	&	1.71	&	16.504	&	125.0	&	4.0	&	24.42	&	4\\
J2234$-$0224	&	22 34 58.76	&	$-$02 24 18.9	&	Q	&	0.5500	&	II	&	3.3	&	1.27	&	18.333	&	75.7	&	2.3	&	25.92	&	3\\
J2239$-$0133	&	22 39 59.34	&	$-$01 33 51.4	&	G	&	0.0881	&	II	&	20.4	&	1.97	&	16.666	&	135.0	&	4.1	&	24.39	&	3\\
J2242$+$6212	&	22 42 32.13	&	$+$62 12 17.6	&	G	&	0.1880	&	II	&	4.2	&	0.78	&	21.896	&	288.7	&	8.8	&	25.44	&	37\\
J2245$-$0032	&	22 45 20.76	&	$-$00 32 06.1	&	G	&	(0.66)	&	II	&	3.0	&	1.25	&	22.018	&	17.0	&	0.6	&	25.46	&	23\\
J2250$+$2844	&	22 50 39.16	&	$+$28 44 45.5	&	G	&	(0.097)	&	II	&	8.4	&	0.93	&	16.747	&	120.0	&	9.0	&	24.44	&	4\\
J2253$-$3455	&	22 53 36.03	&	$-$34 55 30.8	&	G	&	0.2115	&	II	&	4.5	&	0.93	&		&	291.3	&	8.7	&	25.56	&	2\\
J2253$-$5812	&	22 53 58.98	&	$-$58 12 49.4	&	G	&	0.1764	&	II	&	4.3	&	0.76	&		&	5.4	&	0.2	&	23.65	&	12\\
J2256$-$3617	&	22 56 15.08	&	$-$36 17 59.1	&	G	&	0.0902	&	II	&	14.5	&	1.51	&		&	195.0	&	11.0	&	24.58	&	4\\
	&	&	&	&	&	&	&	&	&	&	&	&	\\
J2257$-$0052	&	22 57 34.43	&	$-$00 52 31.8	&	G	&	0.5220	&	II	&	5.3	&	1.98	&	20.790	&	80.2	&	3.0	&	25.89	&	9\\
J2304$-$1050	&	23 04 44.81	&	$-$10 50 47.7	&	G	&	0.2103	&	I	&	4.0	&	0.82	&	17.544	&	67.8	&	2.7	&	24.92	&	4,92\\
J2306$-$0930    &       23 06 32.18	&       $-$09 30 20.6	&	G	&	0.1593  &	I	&       6.2     &       1.01    &	17.519	&	136.9   &	4.5     &	24.96	&	93\\
J2312$+$1356	&	23 12 01.27	&	$+$13 56 55.9	&	G	&	0.1404	&	II	&	11.4	&	1.74	&	16.520	&	338.0	&	26.0	&	25.23	&	4\\
J2312$+$1845	&	23 12 07.57	&	$+$18 45 41.4	&	G	&	0.4265	&	II	&	3.4	&	1.14	&	20.006	&	1957	&	58.7	&	27.08	&	94\\
J2316$-$2823	&	23 16 00.58	&	$-$28 23 52.8	&	G	&	0.2293	&	II	&	8.7	&	1.89	&	17.925	&	296.2	&	15.1	&	25.64	&	2,5\\
J2316$-$0102	&	23 16 20.15	&	$-$01 02 07.3	&	Q	&	(0.221)	&	II	&	6.8	&	1.50	&	19.409	&	90.0	&	5.0	&	25.09	&	4\\
J2320$-$1320	&	23 20 13.98	&	$-$13 20 57.9	&	G	&	0.3927	&	II	&	4.0	&	1.11	&	19.591	&	330.2	&	16.6	&	26.22	&	1\\
J2321$-$2724	&	23 21 13.79	&	$-$27 24 48.6	&	G	&	0.2370	&	II	&	4.7	&	1.06	&	17.731	&	38.9	&	1.2	&	24.79	&	24\\
J2326$+$2458	&	23 26 23.20	&	$+$24 58 40.4	&	G	&	0.2549	&	II	&	11.7	&	2.88	&	18.676	&	274.0	&	27.0	&	25.71	&	4\\
	&	&	&	&	&	&	&	&	&	&	&	&	\\
J2328$-$0825	&	23 28 50.02	&	$-$08 25 11.8	&	G	&	0.3840	&	II	&	4.6	&	1.40	&	19.664	&	64.9	&	3.7	&	25.49	&	1,4\\
J2333$-$2343	&	23 33 55.24	&	$-$23 43 40.7	&	G	&	0.0477	&	II	&	18.9	&	1.05	&	16.321	&	1167	&	35.1	&	24.78	&	2\\
J2335$+$5215	&	23 35 52.14	&	$+$52 15 38.6	&	G	&	0.0706	&	II	&	10.2	&	0.85	&	16.461	&	95.0	&	8.0	&	24.05	&	4\\
J2344$-$0032	&	23 44 40.04	&	$-$00 32 31.7	&	Q	&	0.5014	&	II	&	2.7	&	0.99	&	17.669	&	34.9	&	1.1	&	25.49	&	25\\
J2345$-$0449	&	23 45 32.70	&	$-$04 49 25.4	&	G	&	0.0756	&	II	&	17.1	&	1.46	&	15.968	&	173.9	&	5.2	&	24.38	&	3\\
J2355$+$7955	&	23 55 23.33	&	$+$79 55 19.6	&	G	&	1.3360	&	II	&	1.4	&	0.71	&		&	1722	&	51.7	&	28.20	&	95\\
J2355$+$0256	&	23 55 31.63	&	$+$02 56 07.1	&	G	&	(0.657)	&	II	&	8.1	&	3.48	&	20.619	&	80.0	&	9.0	&	26.13	&	4\\
J2356$-$0131	&	23 56 06.40 	&	$-$01 31 51.4 	&	Q 	&	1.0280 	&	II 	&	1.7	 &	0.81	&	20.820	&	225.6	&	6.8	&	27.02	&	21\\
J2359$-$6054	&	23 59 04.36	&	$-$60 54 59.3	&	G	&	0.0963	&	II	&	~7	&	~0.74	&		&	26240	&	460	&	26.77	&	41\\
	&	&	&	&	&	&	&	&	&	&	&	&	\\
\hline
\end{longtable}
\hspace{-0.63cm}{\bf Column description:} (1) -- source name, (2) and (3) -- J2000.0 coordinates of the host optical object, (4) -- optical identification (G -- galaxy or Q -- quasar), (5) -- redshift, (6) -- radio morphological type, (7) -- angular size in arcmin, (8) projected linear size in Mpc, (9) -- optical aperture magnitude in {\it r} band, (10) total flux-density at 1.4 GHz, unless indicated, in units of mJy , (11) -- error of flux-density in mJy, (12) -- total radio luminosity at 1.4 GHz in units of W$/$Hz (14) -- references.\\
{\bf Notes:} (a) -- flux-density at 1.388 GHz from \cite{saripalli2012}, (b) -- SUMSS measurement of flux-density, (c) -- flux-density at 0.843 GHz from \cite{saripalli2005}, (d) -- flux-density at 0.325 GHz from \cite{sebastian2018}, (r) -- optical {\it r} magnitudes taken from \cite{saripalli2012}. The redshifts given in brackets are photometric, (n) -- in \cite{wezgowiec2016} the diffuse radio emission can be seen. It extends for $\sim$23' and it is presumably the evidence of past radio emission, (m) -- GRG proposed by \cite{mack2005}. Its size is uncertain, (p) -- we also found different value of photometric redshift equal to 0.473 (\citealt{brescia2015}).\\
{\bf References:} 
1. \cite{bankowicz2015},
2. \cite{proctor2016},
3. \cite{machalski2007},
4. \cite{dabhade2017},
5. \cite{solovyov2014},
6. \cite{kapahi1998},
7. \cite{saripalli2012},
8. \cite{subrahmanyan2010},
9. \cite{sadler2007},
10. \cite{saikia2006},
11. \cite{santiago2016},
12. \cite{saripalli2005},
13. \cite{djorgovski1995},
14. \cite{mack1997},
15. \cite{andernach1992},
16. \cite{vanbreugel1982},
17. \cite{laing1983},
18. \cite{subrahmanyan1996},
19. \cite{lacy2000},
20. \cite{colafrancesco2016},
21. \cite{nilsson1998},
22. \cite{fomalont1978},
23. \cite{renteria2017},
24. \cite{sadler2002},
25. \cite{kuzmicz2012},
26. \cite{schoenmakers2001},
27. \cite{tamhane2015},
28. \cite{simpson2006},
29. \cite{koziel2011},
30. \cite{debruyn1989}, 
31. \cite{schoenmakers1998},
32. \cite{saripalli1994},
33. \cite{hunik2016},
34. \cite{reid1999},
35. \cite{hurley2015},
36. \cite{amirkhanyan2015},
37. \cite{lara2001b},
38. \cite{ishwara1999},
39. \cite{filipovic2013},
40. \cite{jagers1986},
41. \cite{malarecki2015},
42. \cite{saripalli1986},
43. \cite{kronberg1986},
44. \cite{cotter1996},
45. \cite{machalski2001},
46. \cite{machalski2006},
47. \cite{hes1995},
48. \cite{kuligowska2009},
49. \cite{lacy1993},
50. \cite{coziol2017},
51. \cite{sebastian2018},
52. \cite{kuligowska2009},
53. \cite{lawgreen1995},
54. \cite{proctor2011},
55. \cite{hogbom1979},
56. \cite{machalski1999},
57. \cite{owen2000},
58. \cite{bhatnagar1998},
59. \cite{kuzmicz2011},
60. \cite{hintzen1983},
61. \cite{vanbreugel1981},
62. \cite{saunders1987},
63. \cite{mack2005},
64. \cite{schoenmakers2000},
65. \cite{machalski1998},
66. \cite{jones1992},
67. \cite{djorgovski1990},
68. \cite{saripalli1997},
69. \cite{mccarthy1997},
70. \cite{parma1996},
71. \cite{gregg2006},
72. \cite{hota2011},
73. \cite{machalski2008},
74. \cite{falco1999},
75. \cite{singal2004},
76. \cite{andernach12},
77. \cite{bassani2016},
78. \cite{masetti2013},
79. \cite{molina2015},
80. \cite{clarke2017},
81. \cite{konar2006},
82. \cite{baum1989},
83. \cite{rottgering1996},
84. \cite{marecki2016},
85. \cite{molina2014},
86. \cite{lacy1999},
87. \cite{lacy1992},
88. \cite{lara1999},
89. \cite{hagen1999},
90. \cite{saripalli2008},
91. \cite{riley1988},
92. \cite{jones2009},
93. \cite{best2005},
94. \cite{leahy1991},
95. \cite{kharb2008}.
                           
\end{landscape}
\end{onecolumn}
\label{lastpage}

\begin{thebibliography}{}
\bibitem[\protect\citeauthoryear{Albareti et al.}{2017}]{Albareti2017}
Albareti, F. D., Allende, P. C., Almeida, A., et al., 2017, ApJS, 233, 25
\bibitem[\protect\citeauthoryear{Amirkhanyan, Afanasiev \& Moiseev}{2015}]{amirkhanyan2015} 
Amirkhanyan, V. R., Afanasiev, V. L., \&  Moiseev, A. V., 2015, AstBu, 70, 45
\bibitem[\protect\citeauthoryear{Andernach et al.}{1992}]{andernach1992} 
Andernach, H., Feretti, L., Giovannini, G., et al., 1992, A\&AS, 93, 331
\bibitem[\protect\citeauthoryear{Andernach et al.}{2012}]{andernach12} 
Andernach, H., Jim\'enez Andrade, E. F., Maldonado S\'anchez, R. F., \&  V\'asquez B\'aez, I. R., 2012, Proceedings of the Science from the Next Generation Imaging and Spectroscopic Surveys Conference, 15-18 October 2012. Online at http://www.eso.org/sci/meetings/2012/surveys2012/posters.html, id. P1
\bibitem[\protect\citeauthoryear{Bankowicz et al.}{2015}]{bankowicz2015} 
Bankowicz, M., Koziel-Wierzbowska, D., \&  Machalski, J., 2015, Proceedings of the SALT Science Conference (SSC2015), 1-5 June 2015. Stellenbosch Institute of Advanced Study. South Africa, Online at http://pos.sissa.it/cgi-bin/reader/conf.cgi?confid=250, id.34
\bibitem[\protect\citeauthoryear{Bassani et al.}{2016}]{bassani2016}
Bassani, L., Venturi, T., Molina, M., et al., 2016, MNRAS, 461, 3165
\bibitem[\protect\citeauthoryear{Baum \& Heckman}{1989}]{baum1989}
Baum, S., A.,  \& Heckman, T., 1989, ApJ, 336, 681
\bibitem[\protect\citeauthoryear{Becker, White \& Helfand}{1995}]{becker1995}
Becker, R.~H., White, R.~L., \& Helfand,D.~J., 1995, ApJ, 450, 559
\bibitem[\protect\citeauthoryear{Best et al.}{2005}]{best2005} 
Best, P. N., Kauffmann, G., Heckman, T. M., \& Ivezi\'c, Z., 2005, MNRAS, 362, 9
\bibitem[\protect\citeauthoryear{Bhatnagar, Krishna \& Wisotzki}{1998}]{bhatnagar1998} 
Bhatnagar, S., Krishna, G., \& Wisotzki, L., 1998, MNRAS, 299, 25
\bibitem[\protect\citeauthoryear{Blundell, \-Raw\-lings \-\& Wil\-lott}{1999}]{blundel1999}
Blundell, K. M., Rawlings, S., \& Willott, C. J., 1999, AJ, 117, 677
\bibitem[\protect\citeauthoryear{Bock, Large \& Sadler}{1999}]{bock1999} 
Bock, D. C.-J., Large, M. I., \& Sadler, E. M., 1999, AJ, 117, 1578 
\bibitem[\protect\citeauthoryear{Brescia, Cavuoti \& Longo}{2015}]{brescia2015}
Brescia, M., Cavuoti, S., \& Longo, G., 2015, MNRAS, 450, 3893
\bibitem[\protect\citeauthoryear{Brown, Webster \& Boyle}{2001}]{brown2001} 
Brown, M. J. I., Webster, R. L., \& Boyle, B. J., 2001, AJ, 121, 2381
\bibitem[\protect\citeauthoryear{Carilli \& Rawlings}{2004}]{carilli2004} 
Carilli, C. L., \& Rawlings, S., 2004, NewAR, 48, 979
\bibitem[\protect\citeauthoryear{Chen et al.}{2011}]{chen2011}
Chen, R., Peng, B., Strom, R. G., \& Wei, J., 2011, MNRAS, 412, 2433
\bibitem[\protect\citeauthoryear{Clarke et al.}{2017}]{clarke2017}
Clarke, A. O., Heald, G., Jarrett, T., et al., 2017, A\&A, 601, 25
\bibitem[\protect\citeauthoryear{Colafrancesco et al.}{2016}]{colafrancesco2016} 
Colafrancesco, S., Mhlahlo, N., Jarrett, T., Oozeer, N., \& Marchegiani, P., 2016, MNRAS, 456, 512
\bibitem[\protect\citeauthoryear{Condon et al.}{1998}]{condon1998} 
Condon, J. J., Cotton, W. D., Greisen, E. W., \& Yin, Q. F., 1998, AJ, 115, 1693
\bibitem[\protect\citeauthoryear{Cotter, Rawlings \& Saunders}{1996}]{cotter1996} 
Cotter, G., Rawlings, S., \& Saunders, R., 1996, MNRAS, 281, 1081
\bibitem[\protect\citeauthoryear{Coziol et al.}{2017}]{coziol2017}
Coziol, R., Andernach, H., Torres-Papaqui, J. P., Ortega-Minakata, R. A., \& Moreno del Rio, F., 2017, MNRAS, 466, 921
\bibitem[\protect\citeauthoryear{Dabhade et al.}{2017}]{dabhade2017} 
Dabhade, P., Gaikwad, M., Bagchi, J., et al., 2017, MNRAS, 469, 2886  
\bibitem[\protect\citeauthoryear{Dawson et al.}{2013}]{dawson2013} 
Dawson, K. S., Schlegel, D. J., Ahn, C. P., et al., 2013, AJ, 145, 10
\bibitem[\protect\citeauthoryear{de Bruyn}{1989}]{debruyn1989} 
de Bruyn, A. G., 1989, A\&A, 226, 13
\bibitem[\protect\citeauthoryear{Djorgovski et al.}{1990}]{djorgovski1990} 
Djorgovski, S., Thompson, D. J, Vigotti, M., \& Grueff, G., 1990, PASP, 102, 113
\bibitem[\protect\citeauthoryear{Djorgovski et al.}{1995}]{djorgovski1995} 
Djorgovski, S. G., Thompson, D. J, Maxfield, L., Vigotti, M., \& Grueff, G., 1995, ApJS, 101, 255
\bibitem[\protect\citeauthoryear{Eales}{1985}]{eales1985}
Eales, S. A., 1985, MNRAS, 213, 899
\bibitem[\protect\citeauthoryear{Eisenstein et al.}{2011}]{eisenstein2011}
Eisenstein, D. J., Weinberg, D. H., Agol, E., et al., 2011, AJ, 142, 72   
\bibitem[\protect\citeauthoryear{Falco et al.}{1999}]{falco1999} 
Falco, E. E., Kurtz, M. J., Geller, M. J., et al., 1999, PASP, 111, 438
\bibitem[\protect\citeauthoryear{Fanaroff \& Riley}{1974}]{fanaroff1974} 
Fanaroff, B. L., \& Riley, J. M., 1974, MNRAS, 167, 31
\bibitem[\protect\citeauthoryear{Filipovic et al.}{2013}]{filipovic2013} 
Filipovic, M. D., Cajko, K. O., Collier, J. D., \& Tothill, N. F. H., 2013, SerAJ, 187, 1
\bibitem[\protect\citeauthoryear{Flewelling et al.}{2013}]{flewelling2016} 
Flewelling, H. A., Magnier, E. A., Chambers, K. C., et al., 2016, 2016arXiv161205243F 
\bibitem[\protect\citeauthoryear{Fomalont et al.}{1978}]{fomalont1978} 
Fomalont, E. B., \& Bridle, A. H., 1978, AJ, 83, 725
\bibitem[\protect\citeauthoryear{Gehrels et al.}{2004}]{gehrels2004} 	
Gehrels, N., Chincarini, G., Giommi, P., et al., 2004, ApJ, 611, 1005
\bibitem[\protect\citeauthoryear{Gregg, Becker \& de Vries}{2006}]{gregg2006} 
Gregg, M. D., Becker, R. H., \& de Vries, W., 2006, ApJ, 641, 210
\bibitem[\protect\citeauthoryear{Gupta et al.}{2017}]{gupta2017} 
Gupta, Y., Ajithkumar, B., Kale, H. S., et al., 2017, Current Science, 113, 707
\bibitem[\protect\citeauthoryear{Hagen, Engels \& Reimers}{1999}]{hagen1999} 
Hagen, H.-J., Engels, D., \& Reimers, D., 1999, A\&AS, 134, 483
\bibitem[\protect\citeauthoryear{Hes, de Vries \& Barthel}{1995}]{hes1995}
Hes, R., de Vries, W. H., \& Barthel, P. D, 1995, A\&A, 299, 17 
\bibitem[\protect\citeauthoryear{Hintzen, Ulvestad \& Owen}{1983}]{hintzen1983} 
Hintzen, P., Ulvestad, J., \& Owen, F., 1983, AJ, 88, 709
\bibitem[\protect\citeauthoryear{Hota et al.}{2011}]{hota2011} 
Hota, A., Sirothia, S. K., Ohyama, Y., et al., 2011, MNRAS, 417L, 36
\bibitem[\protect\citeauthoryear{H\"ogbom}{1979}]{hogbom1979} 
H\"ogbom, J. A., 1979, A\&AS, 36, 173
\bibitem[\protect\citeauthoryear{Hunik \& Jamrozy}{2016}]{hunik2016} 
Hunik, D., \& Jamrozy, M., 2016, ApJ, 817, 1
\bibitem[\protect\citeauthoryear{Hurley-Walker et al.}{2015}]{hurley2015} 
Hurley-Walker, N., Johnston-Hollitt, M., Ekers, R., et al., 2015, MNRAS, 447, 2468
\bibitem[\protect\citeauthoryear{Ishwara\--Chandra \& Saikia}{1999}]{ishwara1999} 
Ishwara\--Chandra, C. H., \& Sai\-kia, D. J., 1999, MNRAS, 309, 100 
\bibitem[\protect\citeauthoryear{J\"agers}{1986}]{jagers1986} 
J\"agers, W.J., 1986, PhD thesis, University of Leiden
\bibitem[\protect\citeauthoryear{Jamrozy et al.}{2008}]{jamrozy2008}
Jamrozy, M., Konar, C., Machalski, J., \& Saikia, D. J., 2008, MNRAS, 385, 1286
\bibitem[\protect\citeauthoryear{Jones \& McAdam}{1992}]{jones1992} 
Jones, P. A., \& McAdam, W. B., 1992, ApJS, 80, 137
\bibitem[\protect\citeauthoryear{Jones et al.}{2009}]{jones2009} 
Jones, D. H., Read, M. A., Saunders, W., et al., 2009, MNRAS, 399, 683
\bibitem[\protect\citeauthoryear{Kapahi}{1989}]{kapahi1989}
Kapahi, V.K. 1989, AJ, 97, 1
\bibitem[\protect\citeauthoryear{Kapahi et al.}{1998}]{kapahi1998} 
Kapahi, V. K., Athreya, R. M., van Breugel, W., McCarthy, P.J., \& Subrahmanya, C.R., 1998, ApJS, 118, 275
\bibitem[\protect\citeauthoryear{Kaiser \& Alexander}{1999}]{kaiser1999}
Kaiser, C. R., \& Alexander, P., 1999, MNRAS, 302, 515
\bibitem[\protect\citeauthoryear{Kaiser, Dennett\--Thorpe \& Alex\-ander}{1997}]{kaiser1997}
Kaiser, C. R., Dennett\--Thorpe, J., \& Alexander, P., 1997, MNRAS, 292, 723
\bibitem[\protect\citeauthoryear{Kharb, O'Dea \& Baum}{2008}]{kharb2008} 
Kharb, P., O'Dea, C. P., \& Baum, S. A., 2008, ApJS, 174, 74
\bibitem[\protect\citeauthoryear{Klein et al.}{2003}]{klein2003} 
Klein, U., Mack, K.-H., Gregorini, L., \& Vigotti, M., 2003, A\&A, 406, 579
\bibitem[\protect\citeauthoryear{Konar et al.}{2004}]{konar2004} 
Konar, C., Saikia, D. J., Ishwara-Chandra, C. H., \& Kulkarni, V. K., 2004, MNRAS, 355, 845
\bibitem[\protect\citeauthoryear{Konar et al.}{2006}]{konar2006} 
Konar, C., Saikia, D.~J., Jamrozy, M., \& Machalski, J., 2006, MNRAS, 372, 693
\bibitem[\protect\citeauthoryear{Koziel-Wierzbowska \& Stasi\'nska}{2011}]{koziel2011} 
Kozie{\l}-Wierzbowska, D., \& Stasi\'nska, G., 2011, MNRAS, 415, 1013
\bibitem[\protect\citeauthoryear{Kronberg, Wielebinski \& Graham}{1986}]{kronberg1986} 
Kronberg, P. P., Wielebinski, R., \& Graham, D. A., 1986, A\&A, 169, 63
\bibitem[\protect\citeauthoryear{Komberg \& Pashchenko}{2009}]{komberg2009} 
Komberg, B. V., \& Pashchenko, I. N., 2009, ARep, 53, 1086
\bibitem[\protect\citeauthoryear{Kuligowska et al.}{2009}]{kuligowska2009} 
Kuligowska, E., Jamrozy, M., Kozie{\l}-Wierzbowska, D., \& Machalski, J., 2009, AcA, 59, 431
\bibitem[\protect\citeauthoryear{Ku\'zmicz, Kuligowska \& Jamrozy}{2011}]{kuzmicz2011} 
Ku{\'z}micz, A., Kuligowska, E., \& Jamrozy, M., 2011, AcA, 61, 71
\bibitem[\protect\citeauthoryear{Ku\'zmicz \& Jamrozy}{2012}]{kuzmicz2012} 
Ku{\'z}micz, A., \& Jamrozy, M., 2012, MNRAS, 426, 851
\bibitem[\protect\citeauthoryear{Lacy, Rawlings \& Warner}{1992}]{lacy1992} 
Lacy, M., Rawlings, S., \& Warner, P. J., 1992, MNRAS, 256, 404
\bibitem[\protect\citeauthoryear{Lacy et al.}{1993}]{lacy1993} 
Lacy, M., Rawlings, S., Saunders, R., \& Warner P. J., 1993, MNRAS, 264, 721
\bibitem[\protect\citeauthoryear{Lacy et al.}{1999}]{lacy1999} 
Lacy, M., Rawlings, S., Hill, G. J., et al., 1999, MNRAS, 308, 1096
\bibitem[\protect\citeauthoryear{Lacy}{2000}]{lacy2000} 
Lacy, M., 2000, ApJ, 536, 1
\bibitem[\protect\citeauthoryear{Laing, Riley \& Longair}{1983}]{laing1983} 
Laing, R. A., Riley, J. M., \& Longair, M. S., 1983, MNRAS, 204, 151
\bibitem[\protect\citeauthoryear{Lara et al.}{1999}]{lara1999} 
Lara, L., M\'{a}rquez, I., Cotton, W. D., et al., 1999, A\&A, 348, 699
\bibitem[\protect\citeauthoryear{Lara et al.}{2000}]{lara2000}
Lara, L., Mack, K. H., Lacy, M., et al., 2000, A\&A, 356, 63
\bibitem[\protect\citeauthoryear{Lara et al.}{2001a}]{lara2001} 
Lara, L., Cotton, W. D., Feretti, L., et al., 2001a, A\&A, 370, 409
\bibitem[\protect\citeauthoryear{Lara et al.}{2001b}]{lara2001b} 
Lara, L., Marquez, I., Cotton, W. D., et al., 2001b, A\&A, 378, 826
\bibitem[\protect\citeauthoryear{Law-Green et al.}{1995}]{lawgreen1995} 
Law-Green, J. D. B., Eales, S. A., Leahy, J. P., Rawlings, S., \& Lacy, M., 1995, MNRAS, 277, 995
\bibitem[\protect\citeauthoryear{Leahy \& Perley}{1991}]{leahy1991} 
Leahy, J. P., \& Perley, R. A., 1991, AJ, 102, 537
\bibitem[\protect\citeauthoryear{Lonsdale et al.}{2009}]{lonsdale2009} 
Lonsdale, C. J., Cappallo, R. J., Morales, M. F., et al., 2009, IEEEP, 97, 1497
\bibitem[\protect\citeauthoryear{Machalski}{1998}]{machalski1998} 
Machalski, J., 1998a, A\&AS, 128, 153
\bibitem[\protect\citeauthoryear{Machalski \& Condon}{1999}]{machalski1999} 
Machalski, J., \& Condon, J. J., 1999, APJS, 123, 41
\bibitem[\protect\citeauthoryear{Ma\-chal\-ski, \-Chy\.zy \& Jamro\-zy}{2004}]{machalski2004} 
Machalski, J., Chy\.zy, K. T., \& Jamrozy, M., 2004, AcA, 54, 391
\bibitem[\protect\citeauthoryear{Machalski, Jamrozy \& Saikia}{2009}]{machalski2009} 
Machalski, J., Jamrozy, M., \& Saikia, D.~J., 2009, MNRAS, 395, 812
\bibitem[\protect\citeauthoryear{Machalski, Jamrozy \& Zo{\l}a}{2001}]{machalski2001} 
Machalski, J., Jamrozy, M., \& Zo{\l}a, S., 2001, A\&A, 371, 445
\bibitem[\protect\citeauthoryear{Machalski et al.}{2006}]{machalski2006} 
Machalski, J., Jamrozy, M., Zo{\l}a, S., \& Kozie{\l}, D., 2006, A\&A, 454, 85
\bibitem[\protect\citeauthoryear{Machalski, Kozie{\l}-Wierzbowska \& Jamrozy}{2007}]{machalski2007} 
Machalski, J., Kozie{\l}-Wierzbowska, D., \& Jamrozy, M., 2007, AcA, 57, 227
\bibitem[\protect\citeauthoryear{Machalski et al.}{2008}]{machalski2008} 
Machalski, J., Kozie{\l}-Wierzbowska, D., Jamrozy, M., \& Saikia, D. J., 2008, ApJ, 679, 149
\bibitem[\protect\citeauthoryear{Machalski et al.}{2011}]{machalski2011} 
Machalski, J., Jamrozy, M., Stawarz, {\L}., \& Kozie{\l}-Wierzbowska, D., 2011, ApJ, 740, 58
\bibitem[\protect\citeauthoryear{Mack et al.}{1997}]{mack1997} 
Mack, K. H., Klein, U., O'Dea, C. P., \& Willis, A. G., 1997, A\&AS, 123, 423
\bibitem[\protect\citeauthoryear{Mack et al.}{1998}]{mack1998} 
Mack, K.-H., Klein, U., O'Dea ,C.~P., Willis, A.~G., \& Saripalli, L., 1998, A\&A, 329, 431
\bibitem[\protect\citeauthoryear{Mack et al.}{2005}]{mack2005} 
Mack, K. H., Vigotti, M., Gregorini, L., et al., 2005, A\&A, 435, 863
\bibitem[\protect\citeauthoryear{Malarecki et al.}{2015}]{malarecki2015} 
Malarecki, J. M., Jones, D. H., Saripalli, L., Staveley-Smith, L., \& Subrahmanyan, R., 2015, MNRAS, 449, 955
\bibitem[\protect\citeauthoryear{Malarecki et al.}{2013}]{malarecki2013} 
Malarecki, J.~M., Staveley-Smith, L., Saripalli, L., et al., 2013, MNRAS, 432, 200
\bibitem[\protect\citeauthoryear{Malkin}{2016}]{malkin2016}
Malkin, Z., 2016, ``A new method to subdivide a spherical surface into equal-area cells'', arXiv:1612.03467
\bibitem[\protect\citeauthoryear{Mao, Berlind \& Scherrer}{2017}]{mao2017} 
Mao, Q., Berlind, A. A., \& Scherrer, R. J., 2017, ApJ, 835, 161 
\bibitem[\protect\citeauthoryear{Marecki, Jamrozy \& Machalski}{2016}]{marecki2016} 
Marecki, A., Jamrozy, M., \& Machalski, J., 2016, MNRAS, 463, 338
\bibitem[\protect\citeauthoryear{Masetti, Parisi \& Palazzi}{2013}]{masetti2013} 
Masetti, N., Parisi, P., \& Palazzi, E., 2013, A\&A, 556, 120
\bibitem[\protect\citeauthoryear{McCarthy et al.}{1997}]{mccarthy1997} 
McCarthy, P. J., Miley, G. K., de Koff, S., et al., 1997, ApJS, 112, 415
\bibitem[\protect\citeauthoryear{Molina et al.}{2014}]{molina2014} 
Molina, M., Bassani, L., Malizia, A., et al., 2014, A\&A, 565, 2
\bibitem[\protect\citeauthoryear{Molina et al.}{2015}]{molina2015} 
Molina, M., Venturi, T., Malizia, A., et al., 2015, MNRAS, 451, 2370
\bibitem[\protect\citeauthoryear{Murgia et al.}{1999}]{murgia1999}
Murgia, M., Fanti, C., Fanti, R., et al., 1999, A\&A, 345, 769
\bibitem[\protect\citeauthoryear{Murgia}{2003}]{murgia2003}
Murgia, M., 2003, Publ. Astron. Soc. Aust., 20, 19
\bibitem[\protect\citeauthoryear{Murgia et al.}{2011}]{murgia2011}
Murgia ,M., Parma, P., Mack, K.-H., et al., 2011, A\&A, 526, 148
\bibitem[\protect\citeauthoryear{Neeser et al.}{1995}]{neeser1995}
Neeser, M. J., Eales, S. A., Law-Green, J. D., Leahy, J. P., \& Rawlings, S., 1995, ApJ, 451, 76
\bibitem[\protect\citeauthoryear{Nilsson}{1998}]{nilsson1998} 
Nilsson, K., 1998, A\&AS, 132, 31
\bibitem[\protect\citeauthoryear{Onah et al.}{2018}]{onah2018}
Onah, C. I., Ubachukwu, A. A., Odo, F. C., \& Onuchukwu, C. C., 2018, RMxAA, 54, 271
\bibitem[\protect\citeauthoryear{Owen et al.}{2000}]{owen2000} 
Owen, F. N, Ledlow, M. J., Eilek, J. A., et al., 2002, IAUS, 199, 171
\bibitem[\protect\citeauthoryear{Parma et al.}{1996}]{parma1996} 
Parma, P., de Ruiter, H. R., Mack, K. H., et al., 1996, A\&A, 311, 49
\bibitem[\protect\citeauthoryear{Parma, Murgia \& Morganti}{1999}]{parma1999}
Parma, P., Murgia, M., \& Morganti, R., 1999, A\&A, 344, 7
\bibitem[\protect\citeauthoryear{Peng, Chen \& Strom}{2015}]{peng2015} 
Peng, B., Chen, R.-R., \& Strom, R., 2015, ``Giant radio galaxies as probes of the ambient WHIM in  the  era  of  the  SKA''  in  proc. Advancing  Astrophysics  with  the  Square  Kilometre  Array, PoS(AASKA14)109, arXiv:1501.00407
\bibitem[\protect\citeauthoryear{Pirya et al.}{2012}]{pirya2012} 
Pirya, A., Saikia, D.~J., Singh, M., \& Chandola, H.~C., 2012, MNRAS, 426, 758
\bibitem[\protect\citeauthoryear{Proctor}{2011}]{proctor2011} 
Proctor, D. D., 2011, ApJS, 194, 31
\bibitem[\protect\citeauthoryear{Proctor}{2016}]{proctor2016} 
Proctor, D. D., 2016, ApJS, 224, 18
\bibitem[\protect\citeauthoryear{Reid, Kronberg \& Perley}{1999}]{reid1999} 
Reid, R. I., Kronberg, P. P., \& Perley, R. A., 1999, ApJS, 124, 285
\bibitem[\protect\citeauthoryear{Rengelink et al.}{1997}]{rengelink1997} 
Rengelink, R.~B., Tang, Y., de Bruyn, A.~G., et al., 1997, A\&AS, 124, 259
\bibitem[\protect\citeauthoryear{Renter\'ia Macario \& Andernach}{2017}]{renteria2017} 
Renter\'ia Macario, J., \& Andernach, H., 2017, arXiv171010731R
\bibitem[\protect\citeauthoryear{Riley et al.}{1989}]{riley1988} 
Riley, J. M. A., Warner, P. J., Rawlings, S., et al., 1989, MNRAS, 236, 13
\bibitem[\protect\citeauthoryear{R\"ottgering et al.}{1996}]{rottgering1996} 
R\"ottgering, H. J. A., Tang, Y., Bremer, M. A. R., et al., 1996, MNRAS, 282, 1033
\bibitem[\protect\citeauthoryear{Sadler et al.}{2002}]{sadler2002} 
Sadler, E. M., Jackson, C. A., Cannon, R. D., et al., 2002,  MNRAS, 329, 227
\bibitem[\protect\citeauthoryear{Sadler, Cannon \& Mauch}{2007}]{sadler2007} 
Sadler, E. M., Cannon, R. D, \& Mauch, T., 2007,  MNRAS, 381, 211
\bibitem[\protect\citeauthoryear{Saikia \& Jamrozy}{2009}]{saikia2009} 
Saikia, D. J., \& Jamrozy, M., 2009, BASI, 37, 63
\bibitem[\protect\citeauthoryear{Saikia, Konar \& Kulkarni}{2006}]{saikia2006} 
Saikia, D. J., Konar, C., \& Kulkarni, V. K., 2006, MNRAS, 366, 139
\bibitem[\protect\citeauthoryear{Santiago-Bautista et al.}{2016}]{santiago2016} 
Santiago-Bautista, I., Rodriguez-Rico, C. A., Andernach, H., et al., 2016, in The Universe of Digital Sky Surveys, Vol. 42, ed. Napolitano N.R., Longo G., Marconi M., Paolillo M., Iodice E., ASSP, 42, 231
\bibitem[\protect\citeauthoryear{Saripalli, Gopal-Krishna \& Kuehr}{1986}]{saripalli1986} 
Saripalli, L., Gopal-Krishna, R. W., \& Kuehr, H., 1986, A\&A, 170, 20
\bibitem[\protect\citeauthoryear{Saripalli, Subrahmanyan \& Hunstead}{1994}]{saripalli1994} 
Saripalli, L., Subrahmanyan, R., \& Hunstead, R. W., 1994, MNRAS, 269, 37
\bibitem[\protect\citeauthoryear{Saripalli et al.}{1997}]{saripalli1997} 
Saripalli, L., Patnaik, A. R., Porcas, R. W., \& Graham, D. A., 1997, A\&A, 328, 78
\bibitem[\protect\citeauthoryear{Saripalli et al.}{2005}]{saripalli2005} 
Saripalli, L., Hunstead, R. W., Subrahmanyan, R., \& Boyce, E., 2005, AJ, 130, 896
\bibitem[\protect\citeauthoryear{Saripalli et al.}{2008}]{saripalli2008} 
Saripalli, L., Subrahmanyan, R., Laskar, T., \& Koekemoer, A., 2008, in {\it From Planets to Dark Energy: 
The Modern Radio Universe}, Proceedings of Science, 130 (astro-ph: 0806.3518)
\bibitem[\protect\citeauthoryear{Saripalli et al.}{2012}]{saripalli2012} 
Saripalli, L., Subrahmanyan, R., Thorat, K., et al., 2012, ApJS, 199, 27
\bibitem[\protect\citeauthoryear{Saunders, Baldwin \& Warmer}{1987}]{saunders1987} 
Saunders, R., Baldwin, J. E., \& Warmer, P. J., 1987, MNRAS, 225, 713
\bibitem[\protect\citeauthoryear{Schoenmakers et al.}{1998}]{schoenmakers1998} 
Schoenmakers, A. P, Mack, K. H., Lara, L., et al., 1998, A\&A, 336, 455
\bibitem[\protect\citeauthoryear{Schoenmakers et al.}{2000a}]{schoenmakers2000} 
Schoenmakers, A.~P., de Bruyn, A.~G., R\"{o}ttgering, H.~J.~A., van der Laan, H., \& Kaiser, C.~R., 2000a, MNRAS, 315, 371
\bibitem[\protect\citeauthoryear{Schoenmakers et al.}{2000b}]{schoenmakers2000b}
Schoenmakers, A. P., Mack, K. H., de Bruyn, A. G., et al., 2000b, A\&AS, 146, 293
\bibitem[\protect\citeauthoryear{Schoenmakers et al.}{2001}]{schoenmakers2001} 
Schoenmakers, A. P, de Bruyn, A. G., R\"ottgering, H. J. A., \& van der Laan, H., 2001, A\&A, 374, 861
\bibitem[\protect\citeauthoryear{Sebastian et al.}{2018}]{sebastian2018}	
Sebastian, B., Ishwara-Chandra, C. H., Joshi, R., \& Wadadekar, Y., 2018, MNRAS, 473, 4926
\bibitem[\protect\citeauthoryear{Simpson et al.}{2006}]{simpson2006} 
Simpson, C., Martinez-Sansigre, A., Rawlings, S., et al., 2006, MNRAS, 372, 741
\bibitem[\protect\citeauthoryear{Singal}{1988}]{singal1988} 
Singal, A. K., 1988, MNRAS, 233, 87
\bibitem[\protect\citeauthoryear{Singal}{1996}]{singal1996} 
Singal, A. K., 1996, IAUS, 175, 563
\bibitem[\protect\citeauthoryear{Singal, Konar \& Saikia}{2004}]{singal2004} 
Singal, A. K., Konar, C., \& Saikia, D. J., 2004, MNRAS, 347, 79
\bibitem[\protect\citeauthoryear{Skrutskie et al.}{2006}]{skrutskie2006} 
Skrutskie, M.F., Cutri, R.M., Stiening, R., et al. 2006, AJ, 131, 1163
\bibitem[\protect\citeauthoryear{Solovyov \& Verkhodanov}{2014 }]{solovyov2014} 
Solovyov, D. I., \& Verkhodanov, O. V., 2014, AstBu, 69, 141
\bibitem[\protect\citeauthoryear{Subrahmanyan, Saripalli \& Hunstead}{1996}]{subrahmanyan1996} 
Subrahmanyan, R., Saripalli, L.,  \& Hunstead, R. W., 1996, MNRAS, 279, 257
\bibitem[\protect\citeauthoryear{Subrahmanyan et al.}{2008}]{subrahmanyan2008}
Subrahmanyan , R.,  Saripalli,  L.,  Safouris,  V.,  \& Hunstead,  R.  W.,  2008,  ApJ, 677, 63
\bibitem[\protect\citeauthoryear{Subrahmanyan et al.}{2010}]{subrahmanyan2010} 
Subrahmanyan, R., Ekers, R. D., Saripalli, L., \& Sadler, E. M., 2010, MNRAS, 402, 2792
\bibitem[\protect\citeauthoryear{Tamhane et al.}{2015}]{tamhane2015} 
Tamhane, P., Wadadekar, Y., Basu, A., et al., 2015, MNRAS, 453, 2438
\bibitem[\protect\citeauthoryear{van Breugel \& Willis}{1981}]{vanbreugel1981} 
van Breugel W. J. M., \& Willis A. G., 1981, A\&A, 96, 332
\bibitem[\protect\citeauthoryear{van Breugel \& J\"agers}{1982}]{vanbreugel1982} 
van Breugel, W. J. M., \& J\"agers, W., 1982, A\&AS, 49, 529
\bibitem[\protect\citeauthoryear{van Haarlem et al.}{2013}]{haarlem2013} 
van Haarlem, M.~P., Wise, M. W.,Gunst, A. W., et al. 2013, A\&A, 556, 2
\bibitem[\protect\citeauthoryear{We\.zgowiec, Jamrozy \& Mack}{2016}]{wezgowiec2016}
We\.zgowiec, M., Jamrozy, M., \& Mack, K.-H., 2016, AcA, 66, 85
\bibitem[\protect\citeauthoryear{Willis, Strom \& Wilson}{1974}]{willis1974}
Willis, A. G., Strom, R. G., \& Wilson, A. S., 1974, Nature, 250, 625
\bibitem[\protect\citeauthoryear{Wing \& Blanton}{2011}]{wing2011}
Wing, J. D., \& Blanton, E. L., 2011, AJ, 141, 88
\bibitem[\protect\citeauthoryear{Winkler}{1994}]{winkler1994}
Winkler, C., 1994, ApJS, 92, 327
\bibitem[\protect\citeauthoryear{Wright et al.}{2010}]{wright2010}
Wright, E. L., Eisenhardt, P. R. M., Mainzer, A. K., et al., 2010, AJ, 140, 1868
\end{thebibliography}
\end{document}